\documentclass[12pt]{article}

\usepackage{epsfig,graphics}
\usepackage{amsmath}
\usepackage{amsfonts}
\usepackage{amssymb}
\usepackage{graphicx}
\usepackage {tabularx} 
\usepackage{rotate}	
\usepackage{slashed}
\usepackage{cite}
\usepackage{pbox}

\newcommand{\bmat}{\left(\begin{array}}
\newcommand{\emat}{\end{array}\right)}

\def\yzero{\smash{\hbox{$y\kern-4pt\raise1pt\hbox{${}^\circ$}$}}}

\def\beq{\begin{equation}}
\def\eeq{\end{equation}}
\def\beqa{\begin{eqnarray}}
\def\eeqa{\end{eqnarray}}

\def\-{\hphantom{-}}

\def\s2{\frac{1}{\sqrt2}}

\def\beq{\begin{equation}}
\def\eeq{\end{equation}}
\def\beqa{\begin{eqnarray}}
\def\eeqa{\end{eqnarray}}

\def\IF{\relax{\rm I\kern-.18em F}}
\def\II{\relax{\rm I\kern-.18em I}}
\def\IP{\relax{\rm I\kern-.18em P}}
\def\IC{\relax\hbox{\kern.25em$\inbar\kern-.3em{\rm C}$}}
\def\IR{\relax{\rm I\kern-.18em R}}

\def\cs{{\cal S}}

\def\Dsl{\,\raise.15ex\hbox{/}\mkern-13.5mu D} 
\def\IZ{Z\kern-.4em  Z}

\def\caja#1{\pbox{20cm}{\rule{0pt}{2.5ex}#1\raise-1.1ex\hbox{\rule{0pt}{2.5ex}}}}

\def\neut{\tilde{\chi}_1^0}
\def\lsim{\raise0.3ex\hbox{$\;<$\kern-0.75em\raise-1.1ex\hbox{$\sim\;$}}}
\def\gsim{\raise0.3ex\hbox{$\;>$\kern-0.75em\raise-1.1ex\hbox{$\sim\;$}}}
\def\bmumu{B_s\rightarrow \mu^+\mu^-}
\def\bsg{b\to s\gamma}

\catcode`\@=11   
\newdimen\@rotdimen
\newbox\@rotbox  

\def\@vspec#1{\special{ps:#1}}
\def\@rotstart#1{\@vspec{gsave currentpoint currentpoint translate
   #1 neg exch neg exch translate}}
\def\@rotfinish{\@vspec{currentpoint grestore moveto}}
\def\@rotr#1{\@rotdimen=\ht#1\advance\@rotdimen by\dp#1%
   \hbox to\@rotdimen{\hskip\ht#1\vbox to\wd#1{\@rotstart{90 rotate}%
   \box#1\vss}\hss}\@rotfinish}
\def\@rotl#1{\@rotdimen=\ht#1\advance\@rotdimen by\dp#1%
   \hbox to\@rotdimen{\vbox to\wd#1{\vskip\wd#1\@rotstart{270 rotate}%
   \box#1\vss}\hss}\@rotfinish}%
\def\@rotu#1{\@rotdimen=\ht#1\advance\@rotdimen by\dp#1%
   \hbox to\wd#1{\hskip\wd#1\vbox to\@rotdimen{\vskip\@rotdimen
   \@rotstart{-1 dup scale}\box#1\vss}\hss}\@rotfinish}%
\def\@rotf#1{\hbox to\wd#1{\hskip\wd#1\@rotstart{-1 1 scale}%
   \box#1\hss}\@rotfinish}%
\def\rotate{\@ifnextchar[{\@rotate}{\@rotate[l]}}
\def\@rotate[#1]#2{\setbox\@rotbox=\hbox{#2}\@nameuse{@rot#1}\@rotbox}

\catcode`\@=12

\topmargin
-1.5cm
\textwidth
15.05cm
\textheight
23.5cm
\oddsidemargin
0.7cm
\evensidemargin
0.7cm

\begin{document}

\makeatletter
\@addtoreset{equation}{section}
\makeatother
\renewcommand{\theequation}{\thesection.\arabic{equation}}
\pagestyle{empty}
\rightline{IFT-UAM/CSIC-12-127} 
\rightline{FTUAM-12-119}
\vspace{0.1cm}
\begin{center}
\LARGE{The NMSSM with F-theory unified \\
boundary conditions
  \\[5mm]}
 \large{L. Aparicio$^{a}$, P.G. C\'amara$^{b,c}$, D.G. Cerde\~no$^{d}$, \\  L.E. Ib\'a\~nez$^{d} $ and I. Valenzuela$^{d}$  \\[6mm]}
\small{
$^a$ International Centre for Theoretical Physics (ICTP), \\[-0.3em]
Strada Costiera 11, I-34014 Trieste, Italy\\
$^b$ Departament d'Estructura i Constituents de la Mat\`eria and Institut de Ci\`encies\\[-0.3em] 
del Cosmos, Universitat de Barcelona, Mart\'{\i} i Franqu\'es 1, 08028 Barcelona, Spain\\
$^c$ Departament de F\'{\i}sica Fonamental,
Universitat de Barcelona, 08028 Barcelona, Spain\\
 $^d$  Departamento de F\'{\i}sica Te\'orica
and Instituto de F\'{\i}sica Te\'orica UAM/CSIC,\\[-0.3em]
Universidad Aut\'onoma de Madrid,
Cantoblanco, 28049 Madrid, Spain 
\\[8mm]}
\small{\bf Abstract} \\[5mm]
\end{center}
\begin{center}
\begin{minipage}[h]{15.0cm}
We study the phenomenological viability of a constrained NMSSM with parameters subject to 
unified boundary conditions from F-theory GUTs.  We find that very simple assumptions about
modulus dominance SUSY breaking in F-theory unification lead to a predictive set of 
boundary conditions, consistent with all phenomenological constraints. The second lightest scalar Higgs $H_2$ can get a mass $m_{H_2} \simeq 125\ \textrm{GeV}$ 
and has properties similar to the SM Higgs. On the other hand 
the lightest scalar $H_1$, with a dominant
singlet component,  would have barely escaped detection at LEP and could be observable at LHC as a peak in $H_1\rightarrow \gamma\gamma$ at around 100 GeV.
 The LSP is mostly singlino and is consistent with WMAP constraints due to coannihilation with the lightest stau, whose mass is in the range $100-250 \ \textrm{GeV}$.
Such light staus may  lead to very characteristic signatures at LHC and be directly searched at linear colliders.  In  these models tan\,$\beta $ is large, of order  $50$,  still the branching ratio for $B_s\rightarrow \mu^+\mu^-$ is consistent with the LHCb bounds and in many cases is also even smaller than the SM prediction. Gluinos and squarks have
masses in the $2 - 3 \ \textrm{TeV}$ region and may  be accessible at the LHC at 14 TeV.  No large enhancement of the $H_2\rightarrow \gamma \gamma $ 
rate over that of the SM Higgs  is expected.

\end{minipage}
\end{center}
\newpage
\setcounter{page}{1}
\pagestyle{plain}
\renewcommand{\thefootnote}{\arabic{footnote}}
\setcounter{footnote}{0}

\section{Introduction}

The starting into operation of LHC is already testing many avenues beyond the 
Standard Model (SM). In particular, the discovery  of a boson with mass around 125~GeV \cite{:2012gu,:2012gk} and  
properties compatible with those of the SM Higgs is significantly constraining many
of these ideas beyond the SM. In this regard one may argue that 
such value for a Higgs mass goes in the direction of 
low energy SUSY, since supersymmetric models predict a lightest Higgs 
with mass $m_h\lesssim 130$ GeV. On the other hand, the observed mass is close to the 
maximum expected in low energy SUSY theories, implying a certain degree of fine-tuning 
in the SUSY-breaking parameters which must be relatively large. This is also consistent with the no observation as yet of  any supersymmetric particle at LHC.

The simplest testing ground for low energy SUSY  is the Minimal Supersymmetric Standard Model (MSSM) which does not involve 
any new particle beyond the SUSY partners required by supersymmetry. Still the MSSM 
has an unattractive ingredient in its bilinear Higgs superpotential term, the $\mu$-term.
Although supersymmetric, this  mass term has to be (for no good reason) of order of the  SUSY-breaking soft terms
to get consistent  electro-weak (EW) symmetry breaking and low energy
SUSY spectrum.  Perhaps the most economical solution to this problem is the 
scale invariant 
Next to Minimal Supersymmetric Standard Model (NMSSM) \cite{Maniatis:2009re,Ellwanger:2009dp} in which a singlet $S$ is added to the MSSM spectrum
and the $\mu$-term is replaced in the superpotential by new  couplings,
\beq
W_{\rm NMSSM} \ =\ W_{\rm Yuk}\ +\ \lambda SH_uH_d \ +\ \frac {\kappa}{3}  S^3 \ .
\eeq
There are no mass parameters in the superpotential and 
the role of the $\mu$-parameter is now played by $\lambda \langle S\rangle$ which upon minimization of
the scalar potential gets naturally of the same order than the SUSY-breaking soft parameters.

The SUSY-breaking 
soft terms involving the singlet $S$ and the Higgs chiral fields have the general form
\beq
V_{soft} ^S\ =\ m_{H_u}^2|H_u|^2 \ +\  m_{H_d}^2|H_d|^2 \ +\ m_{S}^2|S|^2 
\ +\ \left( \lambda A_\lambda SH_uH_d\ +\ \frac {\kappa}{3}A_\kappa S^3 \ +\ h.c. \right) \
\eeq
As in the case of the MSSM, the most general NMSSM model has plenty of free parameters. On the other hand,
in the presence of some underlying unification structure at a high energy scale, one expects the number of parameters to be reduced to a  few. In the case 
of the MSSM, the constrained MSSM (CMSSM) has universal parameters $m$, $M$, $A$, $\mu$ and $B$, where $m$ is the universal scalar mass, $M$ the universal gaugino mass, $A$ the universal trilinear parameter of the standard Yukawa couplings and $B$ the universal bilinear coupling, all of them defined at the unification scale. Indeed, there are 
models based on String  Theory that lead to such universal selection of parameters  or extensions of it (for instance with non-universal Higgs masses), see e.g.~\cite{thebook} and references therein. Similarly, a constrained version of the NMSSM is usually defined in terms of the five universal parameters
\beq
m\ ,\ M\ , \ \lambda \ ,\ \kappa \ , \ A=A_\lambda=A_\kappa\ .
\label{nmssmparam}
\eeq
In practice, however, one usually takes as free parameters $m$, $M$, $A$, tan\,$\beta$ and $\lambda$ (plus the sign of the effective
$\mu$-term). The values of $m_S$ and $\kappa$ are fixed upon minimization and hence $m_S$ in general does not
unify with the rest of the scalars of the theory. The theory is therefore not {\it constrained}  in the same sense as it is in the
CMSSM \cite{Djouadi:2008yj,Djouadi:2008uj}.  One may argue that the singlet  may be  a bit special and is perhaps not surprising that 
$m_S$ is not unified with the rest of the scalar masses. But then it would be inconsistent to unify the trilinear A-terms. 
In particular, $A_\lambda$ and $A_\kappa$ should be unrelated to the Yukawa trilinear coupling $A$. 
Thus, the least one can say is that the  partially constrained versions of the NMSSM considered up to now are slightly inconsistent,
unless one allows $A_\lambda$ and $A_\kappa$ as free parameters, with the resulting reduction of predictivity.
This is one of the main issues that we address in this paper, namely we try to understand and constraint the structure of the NMSSM parameters at a more
 fundamental level.

The NMSSM has received recently a lot of attention after the evidence and subsequent discovery of the  125 GeV boson at LHC \cite{Ellwanger:2011sk,Hall:2011aa,Ellwanger:2011aa,Gunion:2012zd,Arvanitaki:2011ck,King:2012is,Kang:2012sy,Cao:2012fz, Ellwanger:2012ke,Benbrik:2012rm,Gunion:2012gc,Cao:2012yn,Belanger:2012tt,Kowalska:2012gs,King:2012tr}. 
There are two main reasons for that. On the one hand, in a general NMSSM model  the mass of the Higgs particle receives extra 
contributions from the $\lambda SH_uH_d$ superpotential term, 
\beq
m_h^2 \ \simeq \ m_Z^2\ \textrm{cos}^2\,2\beta \ +\ \lambda^2v^2\textrm{sin}^2\,2\beta \ +\  \delta(m_h^2)\label{mh} ,
\eeq
where $v=174$ GeV is the Higgs vev, $m_Z$ is the mass of the $Z$ boson and $\delta(m_h^2)$ denote the loop corrections to the Higgs mass. 
In the MSSM these loop corrections account for the increasing of the Higgs mass from 90 to 125 GeV, requiring 
large soft terms, stop mixing and fine-tuning. In the NMSSM, however, the second term in eq.~(\ref{mh}) gives an
additional contribution to the mass for relatively large $\lambda$ ($\gtrsim 0.5$) and small tan\,$\beta$.
This allows to get a fairly heavy Higgs boson while reducing the fine-tuning.
The second reason for this  recent interest on the NMSSM is the fact that for some regions of the parameter space
one may get an enhanced Higgs decay rate to two photons, as suggested by the   ATLAS and CMS results as of  July 2012 \cite{:2012gu,:2012gk}. 

As we have already mentioned above, our investigation is however not led by these two interests but rather by the attempt to define a fundamentally-motivated  constrained NMSSM and to check it against the present experimental data, including a 125 GeV Higgs. Concretely, as in the CMSSM,  a unified gauge symmetry like $SU(5)$ naturally induces
 universal gaugino masses and also unifies many of the sfermion masses.  Further assumptions about the 
 origin of SUSY-breaking may lead to an increased degree of unification of the SUSY-breaking soft terms. 
 That is for instance the case of modulus dominance SUSY-breaking in F-theory $SU(5)$  GUTs
 (for reviews see ref. \cite{ftheoryreviews}) that we consider here. In refs.~\cite{Aparicio:2008wh,Aparicio:2012iw} boundary conditions of the general form\footnote{Gauge fluxes may slightly distort these boundary condition as we will see
later  \cite{Aparicio:2008wh}.} 
\beq
m^2\, =\, \frac{1}{2}|M|^2\ ,\qquad
A\,  = \, -\frac{3}{2}M \ ,
\label{primerossoft}
\eeq
were phenomenologically analysed in the context of the MSSM. These conditions appear naturally in F-theory $SU(5)$ schemes in which
one assumes that the auxiliary field of the local K\"ahler modulus $T$ is the dominant source of SUSY-breaking, see \cite{Aparicio:2008wh}. In the above references it was found that these boundary conditions are consistent with all low energy constraints, including a Higgs field
with mass around 125 GeV and appropriate  dark matter relic density. 
The scheme is very predictive, implying tan\,$\beta \simeq 41$ and a relatively heavy spectrum with $M\simeq 1.4$ TeV,
leading to squarks and gluinos of order 3 TeV.
Such heavy spectrum is required after imposing the recent bounds on the branching ratio BR($B_s\to \mu^+\mu^-$) from LHCb \cite{Aaij:2012ac} and CMS \cite{Martini:2012np}. 
Actually, the most recent results for this decay \cite{mumucomb,:2012ct} corner very much the parameter space of this very constrained MSSM 
model. 

On the light of the above NMSSM discussion, it is thus natural to explore 
whether the F-theory $SU(5)$ unification idea exploited in \cite{Aparicio:2008wh} for the MSSM may be
extended to the case of the scale invariant NMSSM. In this paper we find  that the boundary conditions (\ref{primerossoft})
are indeed consistent with all the current phenomenological constraints, including a 125 GeV Higgs boson. A large value of tan\,$\beta\simeq 50$ 
is again selected and  a very small $\lambda$ parameter is required. Small values of $\lambda$ naively correspond
to an effective MSSM limit. Still, the presence of the singlet $S$ leads to quite different physics as compared to the MSSM case.
First, the lightest supersymmetric particle is mostly singlino  and correct dark matter abundance is obtained due to coannihilation with the lightest stau,
which is the NLSP, with masses in the range $100-250$ GeV.  Secondly,  in addition to the  SM-like Higgs with mass around 125 GeV, there is 
a lighter neutral scalar with mass around 100 GeV which would have barely escaped detection at LEP and that should be detectable at  LHC. 
Thirdly,  due to an interference effect, the branching ratio BR($B_s\rightarrow \mu^+\mu^-$) is  easily consistent with the recent 
LHCb data and may be even smaller than the SM prediction. Finally, the squark/gluino spectrum is typically lighter than in the MSSM,
with masses as low as 2 TeV, easily accessible at LHC at 13 TeV. All in all, we find that these are highly constrained and predictive scenarios which pass all the current experimental constraints with a scale of SUSY-breaking that can be as low as $M\simeq 850$ GeV.

Another particularly attractive aspect of our theoretical approach is the additional information obtained about the singlet sector.
In the context of modulus dominated SUSY-breaking within local F-theory $SU(5)$
unification, the singlet sector is less determined. Still, in the simplest scenarios we expect additional constraints that, to first approximation, are of the form 
\begin{equation}
A_\lambda\, \simeq \, -M\, \ , \qquad
A_\kappa\, \simeq \,  m_S^2\, \simeq \, 0\ .  
\label{lasotrassoft}
\end{equation} 
Due to the approximate nature of these constraints, we do not impose them in our analysis. However and remarkably, we find
that almost all the region of the parameter space that the current low energy data selects is also consistent with the relationships (\ref{lasotrassoft}) at the unification scale. 
This is particularly remarkable for the linear relationship $A_\lambda \simeq -M$, see figure \ref{fig:linea} in section \ref{sec:consftheory}. Moreover, the geometric structure of 
the F-theory unification predicts small values for the couplings $\lambda$ and $\kappa$, in agreement with the phenomenological requirements. Thus, at least at the semi-quantitative level, combining eqs.~(\ref{primerossoft}) and (\ref{lasotrassoft}), the present NMSSM scheme essentially depends
on only three free parameters, namely $M$, $\lambda$ and $\kappa$, that are reduced (approximately) to two parameters $M$ and $\lambda$ once we impose 
correct EW symmetry breaking. Of course, this is only semi-quantitative since eqs.~(\ref{lasotrassoft}) are only approximate and eqs.~(\ref{primerossoft}) and (\ref{lasotrassoft}) may have small corrections from gauge fluxes in the extra dimensions. However, it explains the very constrained structure of possible 
benchmarks in this scheme, as displayed in table \ref{table:gut} in section \ref{sec:signatures}.

Finally, let us mention that  in this constrained version of the NMSSM we do not find an enhanced Higgs decay rate to $\gamma\gamma$
compared to the SM, but rather a slight reduction. Whereas our numerical analysis takes into account all the known two loop radiative corrections to the Higgs sector of the NMSSM \cite{Degrassi:2009yq}, in this region of the parameter space these corrections are very large. Since some of the two loop corrections to the Higgs sector in the NMSSM have not yet been computed and are thus not taken into account by our numerical analysis, we expect relatively large uncertainties in our numerical estimations for the masses and decay rates of the Higgs sector. Nevertheless, a strong enhancement in this rate, as suggested by the preliminary  ATLAS and CMS recent results \cite{:2012gu,:2012gk}, appears to be clearly disfavoured in this scenario. 

The structure of this paper is as follows. In section \ref{sec:ftheory} we describe the F-theory unification structure in the context of NMSSM models with K\"ahler modulus domination. In section \ref{sec:constr} we run the SUSY-breaking soft parameters from the unification scale down to the EW scale according to the renormalization group equations and perform a computerized scan over the complete parameter space of these models. We also comment on the main phenomenological constraints that shape the allowed regions of the parameter space. The main constraints arise from the Higgs sector, the dark matter relic density and rare decays such as $B_s\to\mu^+\mu^-$. In section \ref{sec:consftheory} we analyze the consistency of the regions of the parameter space that pass all the current phenomenological constraints,  with the F-theory unification structure discussed in section \ref{sec:ftheory}. In section \ref{sec:signatures} we give some benchmarks and discuss their possible signatures at LHC and other future experiments. Finally, we end with some last comments in section \ref{sec:discuss}. All the above analysis takes into account the known two loop radiative corrections to the Higgs sector of the NMSSM. In appendix \ref{app} we repeat the analysis of the Higgs sector but keeping only account of the one loop radiative corrections to this sector, showing still qualitatively good agreement with the results in the main part of the text.

 \section{The NMSSM and F-theory unification}
\label{sec:ftheory}

F-theory  $SU(5)$ unification \cite{ftheoryguts}
has been recently the subject of intense research both from the phenomenological
and string compactification points of view \cite{ftheoryreviews,Aparicio:2012ju}.
These theories provide an ultraviolet completion within string 
theory  to  more traditional $SU(5)$ 
GUTs, while involving new mechanisms to address some of the problems of field theory GUTs. The symmetry breaking 
from $SU(5)$ down to the SM is induced by the generic presence of  hypercharge fluxes in the compact dimensions.
The presence of these fluxes can also give rise to doublet-triplet splitting for appropriate flux choices. 
Furthermore the observed departure of Yukawa couplings for D-quarks and charged leptons from unification
may also be accounted for \cite{Font:2012wq}.
The reader not interested in the formal details of this class of model may safely jump to section \ref{sec:constr}.

F-theory may be considered as a non-perturbative generalization of type IIB orientifold compactifications.
Ten-dimensional type IIB string theory contains a complex scalar dilaton field $\tau=e^{-\phi}+iC_0$, where 
$\phi$ controls the perturbative loop expansion and $C_0$ is the Ramond-Ramond scalar. The theory is invariant under a
$SL(2,{\bf Z})$ symmetry generated by the transformations $\tau\rightarrow 1/\tau$ and $\tau\rightarrow \tau + i$.
F-theory provides a geometrization of this symmetry by adding two (auxiliary) extra dimensions with  $T^2$ toroidal 
geometry and identifying the complex structure of this $T^2$ with the complex dilaton $\tau$. 
The resulting 
geometric construction is 12-dimensional and 
$N=1$ 4d vacua can be obtained by compactifying the theory on a complex Calabi-Yau (CY) 4-fold  $X_4$. Such 4-fold is required to be an {\it eliptic fibration} over a 6-dimensional
base $B_3$, so that locally  $X_4\simeq T^2\times B_3$. 
Codimension-4 singularities of the fibration correspond to 7-branes wrapping 4-cycles of the base $B_3$. In F-theory $SU(5)$ GUTs one set of such 7-branes, wrapping a 4-cycle $\cs$, yield  an $SU(5)$ gauge symmetry.
\begin{figure}[!t]
\hspace*{-0.6cm}
\centering
\includegraphics[width=15.cm, angle=0]{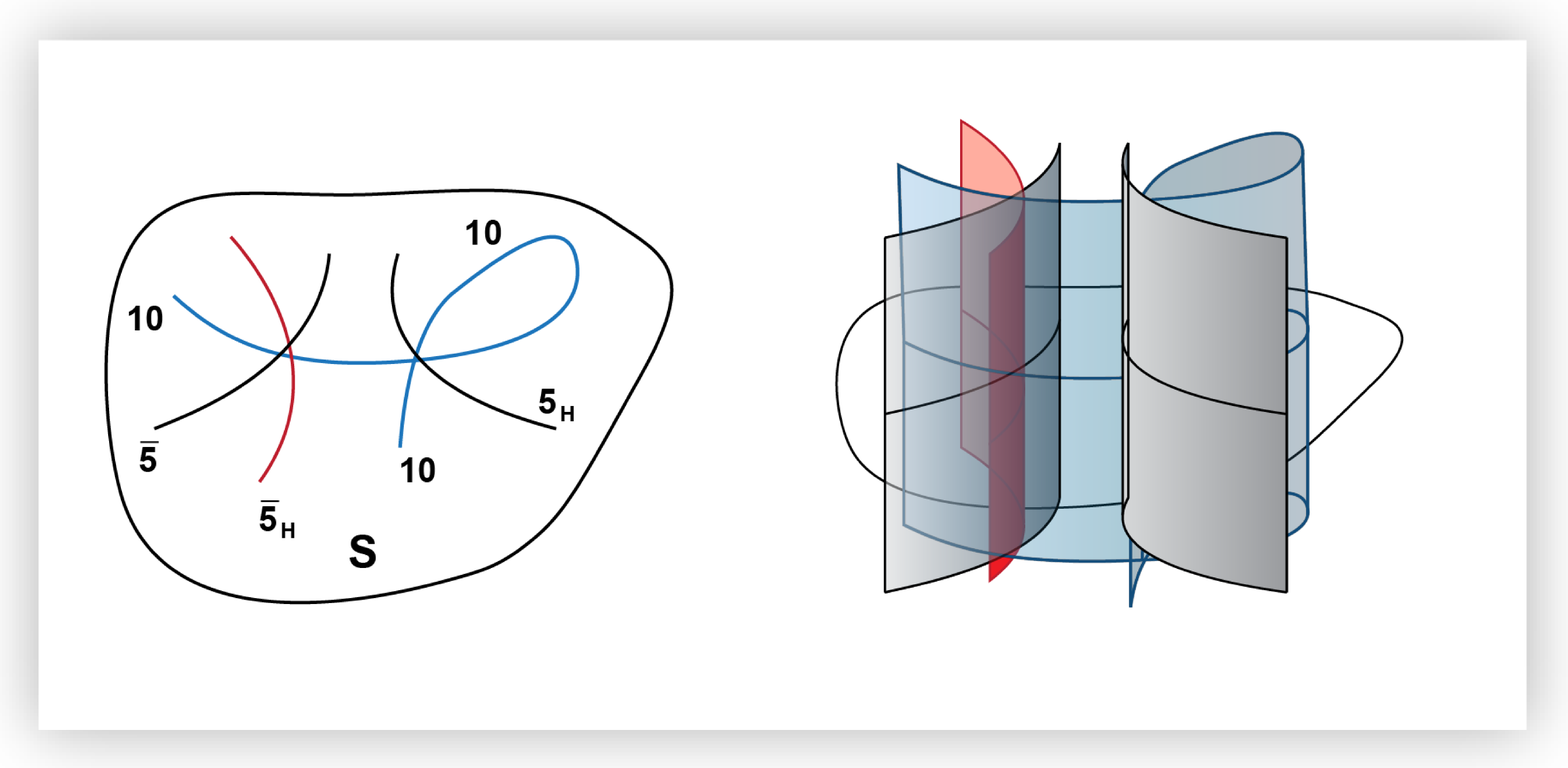}
\caption{General structure of a local  F-theory $SU(5)$ GUT. The GUT group lives on 7-branes whose 4
extra dimensions beyond Minkowski wrap a 4-cycle $\cs$ inside a complex 
3-fold $B_3$, on which the 6 extra dimensions of String Theory are compactified. Gauge bosons live in the bulk of $\cs$ whereas
quarks,  leptons, and Higgsses are localized in complex curves inside $\cs$. These matter curves (denoted as ${\bf 10}$, ${\bf \bar{5}}$, ${\bf 5_H}$ and ${\bf \bar{5}_H}$ in
the figure) correspond to the intersection of 
the 7-branes wrapping $\cs$ with other $U(1)$  7-branes, as depicted in the figure of the right hand side.  There is one matter curve for each $SU(5)$ representation. At the
intersection of matter curves with Higgs curves ${\bf 5_H}$ and ${\bf \bar{5}_H}$, Yukawa couplings develop (figure taken  from ref.~\cite{Aparicio:2012ju}).
\label{ftheoryguts}}
\end{figure} 
Moreover, for suitable topologies of $B_3$, it is possible decouple the local dynamics associated to the $SU(5)$ branes  living on the 4-fold $\cs$ from the global aspects of the
$B_3$ compact space.

Chiral matter appears 
at the complex 1-dimensional pairwise intersections of 7-branes, corresponding to an 
enhanced degree of the singularity (see figure \ref{ftheoryguts}). In F-theory language the locus of the intersection is usually called {\it matter curve}. In minimal $SU(5)$ GUTs the gauge symmetry is locally enhanced to
$SU(6)$ or $SO(10)$  at the matter curves. Indeed, recalling the adjoint branchings
\begin{align}
SU(6)\ & \to \ SU(5)\times U(1) \\
{\bf 35}  & \to\ {\bf 24}_0 \ +\ {\bf 1}_0 \ +\ [ {\bf 5}_1 \ +\ \textrm{c.c.}] \nonumber\\
SO(10) \ & \to \ SU(5)\times U(1)' \\
{\bf 45}  & \to \ {\bf 24}_0 \ +\ {\bf 1}_0 \ +\ [ {\bf 10}_4 \ +\ \textrm{c.c.}] \nonumber  
\end{align}
we observe that in the matter curve associated to a $\bf{5}$ (or a ${\bf \bar{5}}$) representation of $SU(5)$ the gauge symmetry is enhanced to $SU(6)$, whereas
in the one related to a ${\bf 10}$ of $SU(5)$ the gauge symmetry is enhanced to $SO(10)$. In order to get chiral fermions there must be
 non-vanishing fluxes along the $U(1)$ and
$U(1)$' symmetries.
  
In addition to the above matter curves, a third matter curve with an enhanced $SU(6)$' symmetry is required to obtain 
Higgs 5-plets. Yukawa couplings appear at the point-like intersection of the Higgs matter curve with the fermion matter curves, as
illustrated in figure~\ref{ftheoryguts}.
At the intersection point the symmetry is further enhanced to $SO(12)$ in the case of  
${\bf 10}\times {\bf \bar{5}}\times {\bf \bar{5}_H}$ down Yukawa couplings and to $E_6$ in the case of ${\bf 10}\times {\bf 10}\times {\bf 5_H}$ up Yukawa couplings.

\begin{figure}[ht!]
\hspace*{-0.6cm}
\centering
\includegraphics[width=15.cm, angle=0]{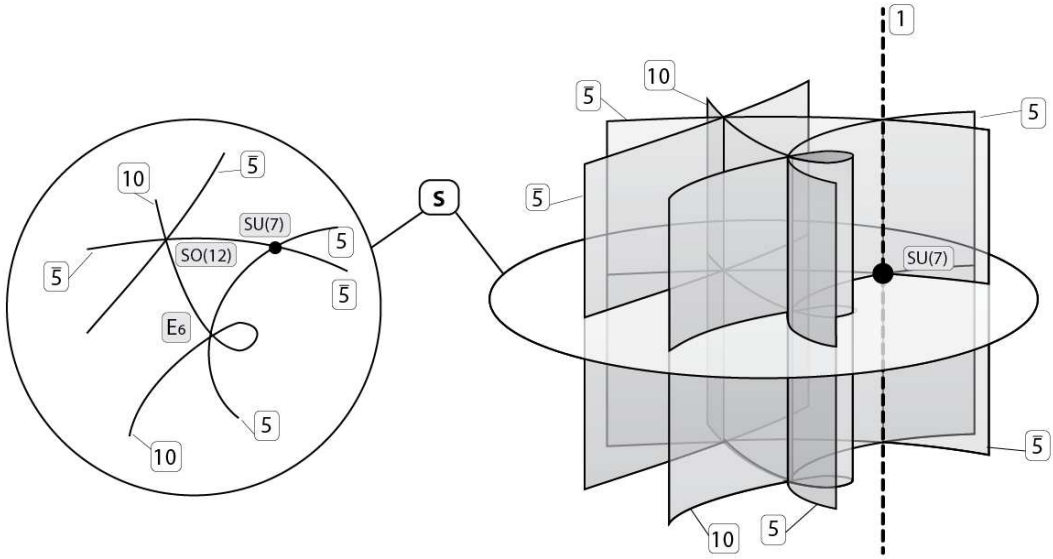}
\caption{General structure of a local  F-theory $SU(5)$ GUT with NMSSM couplings. The matter curves where
the Higgs multiplets live intersect with a transverse curve where the singlet $S$ lives. At the triple intersection point the gauge symmetry is enhanced to $SU(7)$.
\label{ftheorynmssm}}
\end{figure}

In order to make contact with SM physics, the $SU(5)$ gauge symmetry must be broken down to
$SU(3)\times SU(2)\times U(1)$. In these constructions there are no massless adjoints to
make that breaking and discrete Wilson lines are also not available. Nevertheless, one can still achieve such breaking by means of an 
additional  flux $F_Y$ along the hypercharge generator in $SU(5)$, leading to the same symmetry breaking effect
as an adjoint Higgssing. Interestingly enough, this hypercharge flux is also a source for
doublet-triplet splitting of the Higgs multiplets ${\bf 5}_H+{\bf {\bar 5}}_H$.

Besides the above points of $SO(12)$ and $E_6$ enhancement, F-theory $SU(5)$ GUTs may also contain points at which the gauge symmetry is enhanced to $SU(7)$. These correspond to ${\bf 5}\times {\bf \bar{5}}\times 1$ intersections between two 5-plet curves and a singlet curve, as can be derived from the adjoint branching
\begin{align}
SU(7)&\to SU(5)\times U(1)^2\label{su7branch}\\
{\bf 48}& \to  
{\bf 24}+{\bf 1_{(0,0)}}+{\bf 1_{(0,0)}}\ +\ [{\bf 5_{(0,-1)}} + {\bf 1_{(-1,1)}} + {\bf {\bar 5}_{(1,0)}}\ +\ \textrm{c.c.}\ ]\nonumber
\end{align}
The bosons associated to the extra $U(1)$s  are in general  anomalous and become massive
in the usual way.\footnote{F-theory models   with hypercharge flux GUT group breaking and $SU(5)$ singlets charged under $U(1)$ Peccei-Quinn like gauge symmetries have been argued to require the presence of exotics in the spectrum \cite{Marsano:2009wr,Dudas:2010zb,Marsano:2010sq,Palti:2012dd}. Here we assume models with no exotics or with very massive ones. Other constructions, such as those where the singlets are not charged under the extra $U(1)$s (for instance, if they come from closed and/or open string moduli in the microscopic theory) also fit well within this context.}

The simplest models of this class contain one extra singlet $S$ coupling to the Higgs multiplets and  at low energies are equivalent to the scale invariant NMSSM. In what follows we consider that specific setup. The structure of the superpotential is thus given by
\beq
W_{\rm NMSSM} \ =\ W_{\rm Yuk}\ +\ \lambda SH_uH_d \ +\  \frac {\kappa}{3} S^3 \ .
\label{nmssmsuper}
\eeq
with $W_{\rm Yuk}$ the standard superpotential Yukawa couplings. To get this structure the matter curves associated to the two Higgs 5-plets ${H_u}$ and ${H_d}$ must intersect with 
a transverse singlet $S$ curve, see figure \ref{ftheorynmssm}. At the point of intersection the symmetry is enhanced from $SU(6)$ 
 to $SU(7)$. Identifying the singlet in the brackets of eq.~(\ref{su7branch}) with $S$ we see that indeed it may couple to the two Higgs multiplets while respecting the $U(1)$ symmetries. 

In an F-theory construction as above there is a K\"ahler modulus $T_b$ whose real part $t_b$ describes the size 
of the 6-dimensional manifold $B_3$. In addition, the volume of the 4-cycle $\cs$ where the $SU(5)$ degrees of freedom live
is controlled by a local modulus $T$ with real part $t$. Finally, the $U(1)$s live in other 4-cycles different than $\cs$ whose volumes are controlled by K\"ahler moduli that we will denote collectively as $T_S$, with real part $t_S$.
In a local setting one assumes $t,t_S\ll t_b$, so that to leading approximation we can safely ignore the details of $B_3$ in order to study the dynamics of the $SU(5)$ branes.\footnote{The existence of a manifold $B_3$ admitting such a local limit is however not always guaranteed (see e.g. \cite{Ludeling:2011en}).} One can describe this structure by
a {\it Swiss cheese} type K\"ahler potential of the  simplified form \cite{swisscheese}
\beq
K\ =\  -2\textrm{log}\left(t_b^{3/2}\ -\ t^{3/2} \ -\ t_S^{3/2}\right)
\label{kahlerin}
\eeq
where we are working in Planck units. Following \cite{Conlon:2006tj} (see also \cite{thebook}), we can make use of scaling arguments in order to extract the leading moduli dependence of
the K\"ahler metrics of the matter fields.  For a matter field $X$  living on a divisor $\cs_X$ the K\"ahler metric admits an
expansion  for $t_X\ll t_b$   
of the form
\beq
K_X\ =\ \frac { t_X^{(1-\xi_X)}}{t_b}
\eeq
where $\xi_X$ is the so-called {\it modular weight} of the field with respect to the local modulus $t_X$. 
For matter fields localized on a matter curve the mentioned scaling arguments yield $\xi_X=1/2$.
In our case we have respectively for the $SU(5)$ matter multiplets and the singlet $S$
\beq
K_{{\bf \bar{5}},{\bf 10}}\ =\ \frac {t^{1/2}}{t_b} \ ,\qquad  K_S \ =\ \frac {t_S^{1/2}}{t_b}
\label{metricascurvas}
\eeq
In particular, note that the K\"ahler metric of $S$ is independent of $t$. This fact will be relevant for the structure
of soft terms discussed below.

The values of $t_b$ and $t$ are related to known 
physical quantities. In particular in the perturbative regime one obtains (see e.g.\cite{thebook})
\beq
t_b\ =\ \frac {1}{4}  \left(  \frac { M_p}{M_s}\right)^{4/3}  g_s^{1/3} \ \ ,\ \qquad
t\ =\ \alpha_G^{-1}\ \simeq \ 24
\eeq
where $\alpha_G\simeq 1/24$ is the gauge coupling at the unification scale, $M_s=(\alpha')^{-1/2}$ is the string scale and $g_s=\langle e^\phi\rangle$ the string coupling constant.
For $M_s\simeq 10^{16}$ GeV one has $t_b\simeq 115\, g_s^{1/3}\gg  t$ for $g_s\gg 0.01$, consistently with our requirement $t\ll t_b$.

\subsection{The couplings $\lambda$ and $\kappa$}
\label{sec:lambdakap}

Due to the above hierarchy of volumes, physical Yukawa couplings (namely, those that correspond to normalized kinetic terms) may differ 
substantially from the holomorphic Yukawas appearing in the superpotential. Physical Yukawa couplings $h_{\rm phys}$ are related to the holomorphic ones $h_0$ 
by the standard $N=1$ supergravity expression
\beq
h_{\rm phys}^{ijk} \ =\ h_{0}^{ijk}(K_iK_jK_k)^{-1/2}e^{K/2}
\eeq
where $K_i$ are the (diagonal) K\"ahler metrics of the fields involved in the coupling and we do not sum over $i$, $j$ and $k$. In particular, for the 
NMSSM specific Yukawa couplings in eq.~(\ref{nmssmsuper}) we obtain
\begin{align}
\lambda_{\rm phys}  &=  \lambda_0 \,  t^{-1/2}\, t_S^{-1/4}\ \simeq  \ \lambda_0 \, \frac  { \alpha_G^{1/2}}{t_S^{1/4}} \, ,\\
\kappa_{\rm phys}  &=  \kappa_0  \,  t_S^{-3/4} \, . \nonumber
\end{align}
where we have made use of eqs.~(\ref{kahlerin}) and (\ref{metricascurvas}). 
For $t_S \simeq t \simeq 24$ we thus get a suppression factor in relating the holomorphic
couplings $\lambda_0$ and $\kappa_0$ to the physical ones, with $\lambda_{\rm phys}, \kappa_{\rm phys} \simeq 9\times 10^{-2}\, \lambda_0,\kappa_0$. We will show in section \ref{sec:constr} that appropriate EW symmetry breaking in these models together with the current experimental constraints require  small  physical couplings $\lambda_{\rm phys}$ and $\kappa_{\rm phys}$. This is indeed consistent with the above suppression coming from the K\"ahler metrics, as typically $|\lambda_0|, |\kappa_0|\leq 1$. On the contrary, large values for these couplings as required e.g. in refs.~\cite{Ellwanger:2011sk,Hall:2011aa,Ellwanger:2011aa,Arvanitaki:2011ck,King:2012is,Kang:2012sy,Cao:2012fz, Ellwanger:2012ke,Gunion:2012gc,Cao:2012yn,Belanger:2012tt,King:2012tr} do not seem viable within our context.

We have already mentioned in the previous subsection the possible origin of the cubic $\lambda$ coupling in eq.~(\ref{nmssmsuper}) from the intersection of three matter curves on a point of enhanced $SU(7)$ symmetry. If that is the case, the cubic $\kappa$ coupling in eq.~(\ref{nmssmsuper}) should arise from
instanton corrections that violate the (typically anomalous) $U(1)$s, under which $S$ is charged (see e.g. \cite{Cvetic:2010dz}). In particular, the absence of a quadratic $S^2$ coupling suggests the presence of a ${\bf Z}_3$ 
gauged symmetry remnant of a gauged $U(1)_X$ symmetry \cite{berasaluce}.  
One thus expects  $\kappa$ to be a small parameter.\footnote{Cubic self-couplings could also arise if $S$ were in fact a linear combination 
of three different singlets within an extended  gauge symmetry $E_7$ or $E_8$ in F-theory. Even in this case a small 
coupling $\kappa$ is expected due to the geometric suppression discussed above.}

\subsection{$T$-modulus dominance and unified soft-terms}

A natural source of supersymmetry breaking in type IIB string theory compactifications, or more generally in F-theory, is the presence of certain classes of closed string antisymmetric fluxes ($G_4$ fluxes in F-theory). As described in \cite{softfromflux}, from the point of view of the 4-dimensional effective supergravity, the supersymmetry breaking fluxes are encoded in non-zero vevs of the F-auxiliary fields of the K\"ahler moduli. In this work we assume a hierarchy of vevs
$F_{t_b}\gg F_{t}\gg F_{t_S}$ as it will lead to a very constrained set of soft-terms. Such hierarchies of auxiliary fields easily arise in large volume models (see \cite{swisscheese}).
Thus, following \cite{Aparicio:2008wh}  and making use of eqs.~(\ref{kahlerin}) and (\ref{metricascurvas}) we can compute explicitly the
structure of soft terms at the unification scale for the MSSM sector of the theory, yielding results as in \cite{Aparicio:2008wh}, namely
\begin{align}
M\, &=\, \frac{F_t}{t}\ ,
\label{fgut}\\
m_{H}^2\, &=\, \frac{|M|^2}{2}\left(1-\frac32\rho_H\right)\ ,\nonumber\\
m_{{\bf \bar{5}},{\bf 10}}^2\, & =\, \frac{|M|^2}{2}\ ,\nonumber\\
A\, &= \, -\frac{M}{2}(3-\rho_H)\nonumber  \ .
\end{align}
where $M$ are universal gaugino masses, $A$ are trilinear couplings (which appear multiplied by the
SM Yukawa couplings in the Lagrangian) and $m_{H}$, $m_{\bf \bar{5}}$ and $m_{\bf 10}$ are universal  masses for the scalar fields in the three SM matter curves. 
The parameter $\rho_H$ corresponds to a correction describing the effect of 
gauge fluxes on the Higgs matter curve, see \cite{Aparicio:2008wh} for details. This parameter should be small, of order 
$\rho_H\simeq \alpha_G^{1/2}\simeq 0.2$ or smaller.

In addition to the above MSSM-like soft terms, there are also soft terms that involve the singlet $S$. The structure of such soft terms is however more subtle since, unlike the SM fields and gauge bosons, 
$S$ is not localized on the 4-cycle $\cs$ and therefore may be subject to extra sources of supersymmetry breaking. 
The statements that one can make for the soft terms involving  $S$ are thus more model dependent. 
In the simplest case, with no other sources of supersymmetry breaking  for $S$ other than 
$F_{t_b}$ and $F_t$, making use of the K\"ahler metric for $S$, eq.~(\ref{metricascurvas}), yields 
at the unification scale
\begin{equation}
A_\lambda\, = \, -M\,(1-\rho_H)\ , \qquad
A_\kappa\, = \,  m_S^2\, = \, 0\ , \label{singletgut}
\end{equation}
where $A_{\lambda}$ and $A_\kappa$ are the coefficients of the trilinear scalar couplings $ \lambda (SH_uH_d)$ and   $\kappa S^3$ in the Lagrangian.\footnote{By a slight abuse of notation, we make use of the same letter to denote the multiplet (appearing in the superpotential) and its scalar component (appearing in the Lagragian).} 

In the next section we analyze the phenomenological implications of the boundary conditions (\ref{fgut}) that depend on the two
parameters $M$ and $\rho_H$, and we leave $\lambda$, $\kappa$, $A_\lambda$, $A_\kappa$ and $m_S$ as free parameters.  
However,
as described in detail in section \ref{sec:consftheory}, the regions of the parameter space  that are consistent with correct
EW symmetry breaking, a  Higgs mass in the vicinity of 125 GeV and other phenomenological constraints
tend to select solutions that are also consistent with the constraints (\ref{singletgut})! This is remarkable since in no place such second stronger unification conditions were imposed.

\section{Higgs masses, dark matter and other constraints}
\label{sec:constr}

In order to analyze the phenomenological prospects of the NMSSM models with F-theory unification boundary conditions discussed above, in this section we run the soft terms eqs.~(\ref{fgut}) from the unification scale down to the EW scale according to the renormalization group equations, and impose radiative EW symmetry breaking in the standard way as well as the main phenomenological constraints. The strongest phenomenological constraints that shape the structure of our solutions come from the Higgs sector, the dark matter relic density and the branching ratio BR($B_s\to\mu^+\mu^-$), as we discuss below.
 
\subsection{Scanning over modulus dominated NMSSM vacua}
\label{subsec:scan}

Without imposing the extra conditions (\ref{singletgut}), NMSSM models with F-theory unification boundary conditions consist of seven parameters, namely the values of $M$, $\rho_H$, $\lambda$, $\kappa$, $A_\lambda$, $A_\kappa$ and $m_S$ at the GUT scale. One combination of these parameters is however fixed by requiring the correct pole mass for the $Z$ boson, $m_Z=91.187$~GeV \cite{Beringer:1900zz}. Equivalently, we can take the six independent parameters to be given by the values of $M$, $\rho_H$, $A_\lambda$ and $A_\kappa$ at the unification scale, $\lambda$ at the supersymmetry breaking scale and tan\,$\beta$ at the scale $m_Z$. We perform a large scan over this parameter space by means of a modified version of the computer program \texttt{NMSSMTools} \cite{Ellwanger:2004xm} v3.2.0, where we have implemented the boundary conditions eqs.~(\ref{fgut}) at the unification scale. We take the latest combined result from Tevatron  for the top quark pole mass, $m_t=173.2$ GeV \cite{Aaltonen:2012ra}. 

The output of \texttt{NMSSMTools} consists of the supersymmetric mass spectrum computed at the scale $Q^2=m_{\tilde q_3}m_{\tilde u_3}$, with ${\tilde q_3}$ and ${\tilde u_3}$ the third generation squarks; as well as the main couplings and reduced cross sections at the supersymmetry breaking scale, taking into account the known two-loop radiative corrections to the NMSSM Higgs sector \cite{Degrassi:2009yq}. Moreover, it imposes the main experimental constraints from LEP, Tevatron, LHC and WMAP, that we have updated  with the most recent CMS results for the observed 95\% confidence level upper limit on the reduced cross sections for the Higgs decays $H\to \tau^+\tau^-$ \cite{htautau}, $H\to W^+W^-\to \ell^+\ell^-\nu\nu$ \cite{hww}, $H\to ZZ \to \ell^+\ell^+\ell^-\ell^-$ \cite{hzz} and $VH\to Vb\bar b$ \cite{hbb}, based on 17 fb${}^{-1}$ of integrated luminosity at $\sqrt{s}=7-8$ TeV, the ATLAS and CMS recent results for $H\to \gamma\gamma$ \cite{cmsgamma,atlasgamma}, based on near 11 fb${}^{-1}$ of integrated luminosity at $\sqrt{s}=7-8$ TeV, and the ATLAS result for the MSSM decay $H^+\to \tau^+\nu$ \cite{Aad:2012tj}, based on 4.6 fb${}^{-1}$ of integrated luminosity at $\sqrt{s}=7$ TeV. We have also included the recent LHCb result measuring BR$(B_s\to\mu^+\mu^-)=3.2{+1.5 \atop -1.2}\times 10^{-9}$ at 95\% confidence level \cite{:2012ct}. Regarding the relic density of neutralino dark matter, \texttt{NMSSMTools} calls the computer program \texttt{micrOMEGAs} \cite{Belanger:2001fz} v2.2 in order to estimate it and compares with the amount of cold dark matter observed by the WMAP satellite, $0.1008 \leq \Omega h^2\leq 0.1232$ at the $2\sigma$ confidence level \cite{Komatsu:2010fb}. 
We do not impose a priori  any constraint on the muon anomalous magnetic moment $(g-2)_\mu$ but will comment on the
numerical results for this quantity below.

\begin{table}[!t] \footnotesize
\renewcommand{\arraystretch}{1.50}
\begin{center}
\begin{tabular}{|c||c|c|c|c|c|c|c|c|}
 \hline
 Point & M & $\tan\beta$ & $\rho_H$ & $\lambda^{GUT}$ & $\kappa^{GUT}$ & $A_{\lambda}^{GUT}$ & $A_{\kappa}^{GUT}$ & $m_{S}^{GUT}$      \\
\hline\hline
 $P_{1}$ & 858.2 & 47.5 & 0.035 & $5.2\cdotp10^{-3}$ & $4.1\cdotp10^{-4}$ & -895.4 & -319.2  &  105.9   \\
 \hline
 $P_{2}$ & 888.0 & 49.0 & 0.0092 & $6.5\cdotp10^{-3}$ & $3.8\cdotp10^{-4}$ & -922.9 & -195.6  &  50.0   \\
 \hline
 $P_{3}$ & 981.4 & 49.5 & 0.021 & $6.7\cdotp10^{-3}$ & $4.1\cdotp10^{-4}$ & -1015.7 & -262.2  &  82.9   \\
 \hline
 $P_{4}$ & 1009.9 & 52.9 & 0.195 & $6.4\cdotp10^{-3}$ & $3.7\cdotp10^{-4}$ & -1045.7 & -252.7  &  78.5   \\
 \hline
 $P_{5}$ & 1036.0 & 51.5 & 0.135 & $6.3\cdotp10^{-3}$ & $3.9\cdotp10^{-4}$ & -1070.3 & -303.1  &  99.9   \\
 \hline
 $P_{6}$ & 1180.5 & 53.3 & 0.030 & $5.5\cdotp10^{-3}$ & $2.5\cdotp10^{-4}$ & -1204.2 & -198.2  &  50.6   \\
 \hline
$P_{7}$ & 1218.5 & 56.8 & 0.30 & $5.7\cdotp10^{-3}$ & $2.8\cdotp10^{-4}$ & -1257.9 & -281.6  & 91.9   \\
 \hline
$P_{8}$ & 1271.0 & 55.1 & 0.182 & $5.1\cdotp10^{-3}$ & $1.9\cdotp10^{-4}$ & -1290.0 & -160.5  &  22.7   \\
 \hline
 \end{tabular}
\end{center} \caption{F-theory unification/modulus dominance
 boundary conditions  for a representative of benchmark points satisfying all the experimental constraints. Masses are given in GeV.\label{table:gut}}
\end{table}

\begin{table}[!t] \footnotesize
\renewcommand{\arraystretch}{1.50}
\begin{center}
\begin{tabular}{|c||c|c|c|c|c|c|}
\hline 
Point & $\lambda^{SUSY}$ & $\kappa^{SUSY}$ & $A_{\lambda}^{SUSY}$ & $A_{\kappa}^{SUSY}$ & $m_{S}^{SUSY}$ & $\mu_{eff}$  \\  
\hline\hline
  $P_{1}$  & $4.8\cdotp10^{-3}$ & $4.1\cdotp10^{-4}$ & -107.0 & -319.1  &  105.8 & 1243  \\
 \hline
 $P_{2}$  & $5.9\cdotp10^{-3}$ & $3.8\cdotp10^{-4}$ & -84.3 & -195.5  &  49.9 & 1280  \\
 \hline 
  $P_{3}$ &  $6.1\cdotp10^{-3}$ & $4.1\cdotp10^{-4}$ & -97.2 & -262.2  &  82.8 & 1401  \\
\hline
 $P_{4}$ &  $5.7\cdotp10^{-3}$ & $3.7\cdotp10^{-4}$ & -98.9 & -252.6  &  78.4 & 1447  \\
 \hline 
 $P_{5}$ &  $5.7\cdotp10^{-3}$ & $4.0\cdotp10^{-4}$ & -107.3 & -303  &  99.9 & 1478  \\
 \hline 
 $P_{6}$ &  $4.9\cdotp10^{-3}$ & $2.5\cdotp10^{-4}$ & -91.0 & -198.2  &  50.5 & 1662  \\
 \hline
$P_{7}$ &  $4.9\cdotp10^{-3}$ & $2.8\cdotp10^{-4}$ & -107.3 & -281.5  &  91.9 & 1723  \\
 \hline
$P_{8}$ &  $4.5\cdotp10^{-3}$ & $1.9\cdotp10^{-4}$ & -86.0 & -160.5  &  22.5 & 1780  \\
 \hline
 \end{tabular}
\end{center} \caption{Boundary conditions at SUSY scale. Masses are given in GeV.}
\label{table:susy}
\end{table}

\begin{figure}[!t]
\begin{center}
\vspace{0.3cm}
\includegraphics[width=15cm]{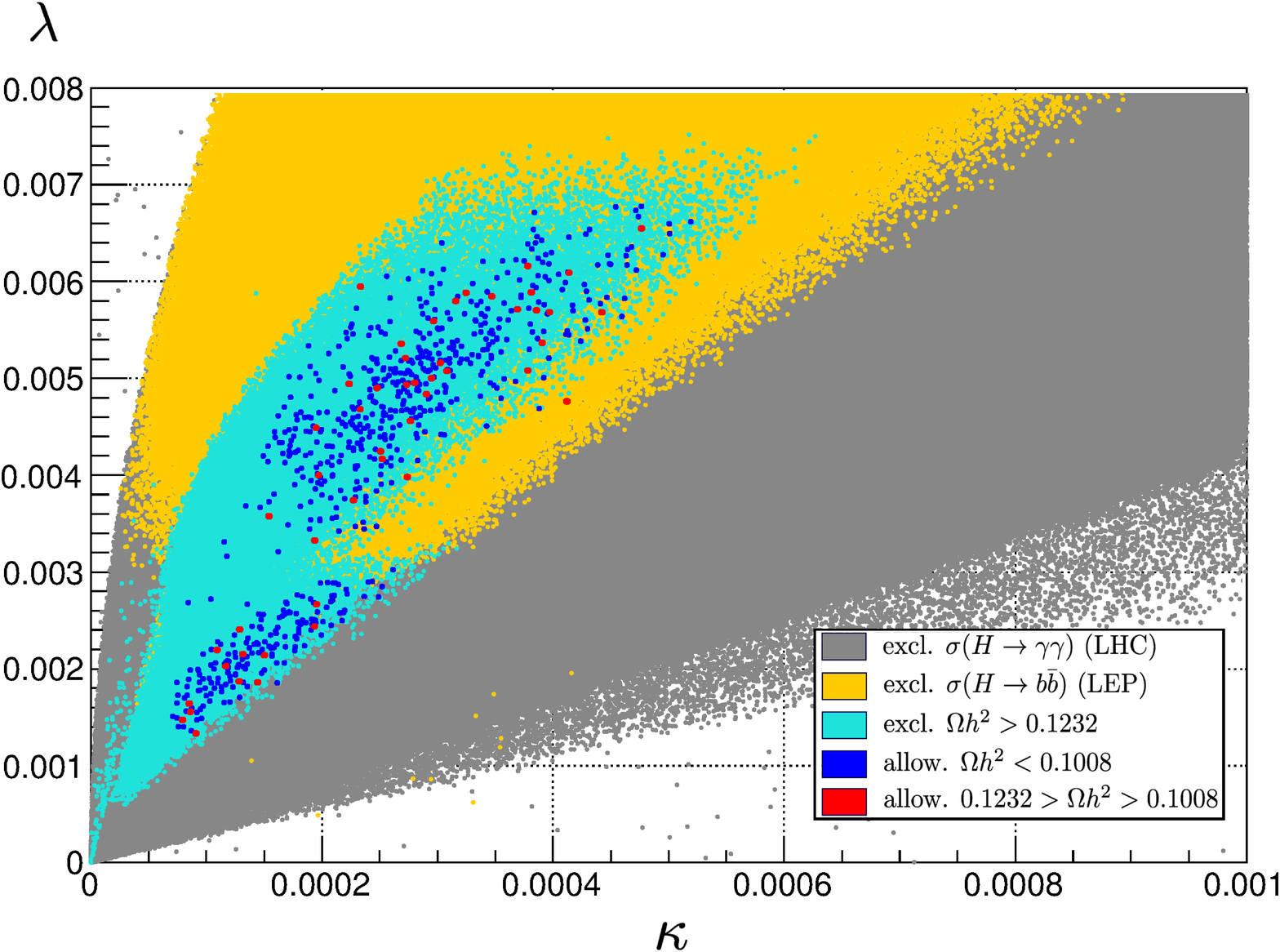}
\caption{\small Distribution of NMSSM vacua over the plane $(\lambda,\, \kappa)$ with F-theory boundary conditions at the unification scale given by eqs.~(\ref{fgut}) and unconstrained values of $A_\lambda$, $A_\kappa$ and $\lambda$. Dark blue and red points both pass all the current experimental and theoretical constraints. Dark blue points, however, have a deficiency of neutralino relic density and therefore require some additional source of dark matter.\label{fig:lambdakappa}}
\end{center}
\end{figure}

Despite the large amount of constraints, we find solutions with correct EW symmetry breaking that pass all the current theoretical and phenomenological constraints and have a moderately low scale of supersymmetry breaking, $M\gtrsim 850$ GeV. All these vacua have large tan\,$\beta$ and small $\lambda$, typically with tan\,$\beta\gtrsim 46$ and $\lambda \lesssim 0.01$ (see figure \ref{fig:lambdakappa} below and figure \ref{fig:m12tanb} in section \ref{sec:mrho}). 
The input parameters for a representative sample of benchmark points are shown in table \ref{table:gut}, where we also include $\kappa^{GUT}$ and $m_S^{GUT}$. The resulting values of the NMSSM parameters at the SUSY scale are indicated in table \ref{table:susy}.
Having such large values of tan\,$\beta$ introduces important uncertainties in the computation of the masses for the Higgs sector, as threshold corrections to down Yukawa couplings become large in this regime (see e.g. \cite{Ross:2007az,Elor:2012ig} for similar considerations in the context of the MSSM). Moreover, in this region of the parameter space also two-loop radiative corrections to the neutral scalar and pseudoscalar masses are large and particularly sensitive to the value of $m_t$. Whereas most of these radiative corrections have been computed \cite{Degrassi:2009yq} and are actually taken into account by \texttt{NMSSMTools}, some of them remain yet unknown. The reader should hence bear in mind that some of our numerical estimations (mostly those concerning the masses of the neutral scalar and pseudoscalar sectors) may contain relatively large uncertainties\footnote{A similar statement is obtained from comparison with the results obtained by means of the \texttt{SARAH-Spheno} \cite{Staub:2008uz} NMSSM implementation presented in \cite{Staub:2010ty} that includes the two-loop radiative corrections of \cite{Degrassi:2009yq}.}. Nevertheless, we find that the qualitative conclusions of our analysis are solid, and do not change if the analysis is done for instance at the one loop level (see appendix \ref{app}).

On the other hand, having small values of $\lambda$ is consistent with the F-theory considerations discussed in section \ref{sec:lambdakap}, and drives these solutions near the effective MSSM limit of the NMSSM. In spite of  the apparent similitude with the MSSM, there are however important differences, coming from the fact that for such small values of $\lambda$ the lightest scalar and pseudoscalar Higgsses and the lightest neutralino are largely dominated by their singlet and singlino components, respectively. This has important consequences for the neutralino relic density and the Higgs couplings, as we discuss below. 

Whereas we have scanned for arbitrary signs and values of $A_\lambda$, $A_\kappa$ and $\lambda$, we have only found phenomenologically viable vacua for the sign choice $A_\lambda<0$, $A_\kappa\leq 0$ and $\lambda\geq 0$. We will discuss further this point in section \ref{subsec:singlet}. In what follows we focus in that region. We have represented in figure \ref{fig:lambdakappa} the values of $\lambda$ and $\kappa$ at the scale of supersymmetry breaking for the set of NMSSM vacua considered in this paper. As we observe, for $\lambda>0$ we recover positive values of $\kappa$ and generally one order of magnitude smaller, $\kappa\lsim\lambda/10$. This is a consequence of the minimization conditions of the scalar potential and is consistent with what happens in the constrained NMSSM \cite{Djouadi:2008uj}.

\subsection{The Higgs sector at two loops}
\label{subsec:higgs}

In the NMSSM the Higgs sector consists of three neutral scalars, $H_{i}$, $i=1,2,3$, two neutral pseudoscalars, $A_{k}$, $k=1,2$, and one charged scalar, $H^+$. Scanning over modulus dominated NMSSM vacua as described in the previous subsection, reveals that the boundary conditions (\ref{fgut}) do not allow for $m_{H_1}\gtrsim 120$ GeV in any region of the parameter space. The only possibility is hence that $H_2$ plays the role of the SM-like Higgs boson, with $m_{H_2}>m_{H_1}$, and $H_1$ is fairly decoupled from the SM sector such that it escaped detection at LEP. This leads naturally to small values of $\lambda$ since in that case $H_1$ is mostly singlet (see figure \ref{fig:lambdakappa}) and it couples very weakly to the SM fields. Thus, one of the main predictions of NMSSM models with F-theory unified boundary conditions (\ref{fgut}) is the presence of an extra light Higgs that is fairly decoupled from the SM fields. Such possibility has also been recently considered in \cite{Belanger:2012tt}.

\begin{figure}[!t]
\begin{center}
\vspace{0.3cm}
\includegraphics[width=15.2cm]{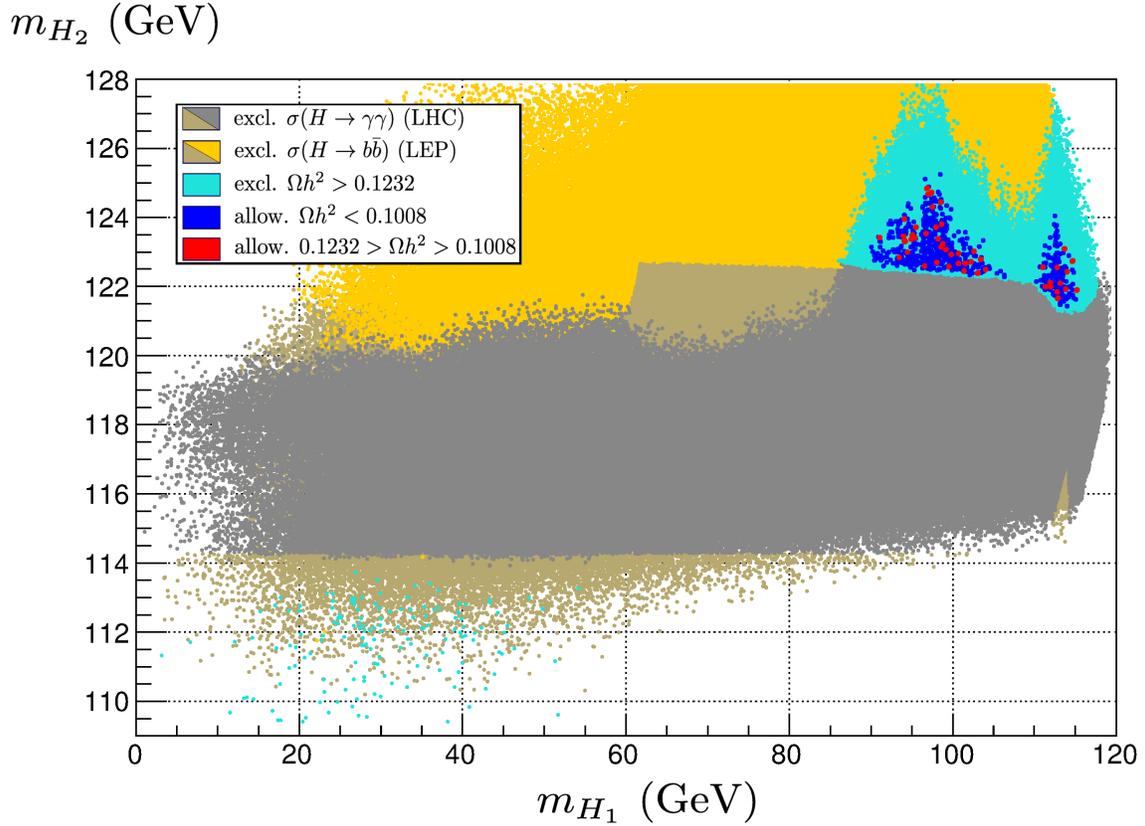}
\caption{\small Distribution of the light Higgs masses $(m_{H_1},m_{H_2})$ for NMSSM vacua with F-theory unification boundary conditions (\ref{fgut}) and unconstrained values of $A_\lambda$, $A_\kappa$ and $\lambda$. Dark blue and red points both pass all the current experimental and theoretical constraints. Dark blue points, however, have a deficiency of neutralino relic density and therefore require some additional source of dark matter.\label{fig:h0h1}}
\end{center}
\end{figure}

To be more precise, we represent in figure \ref{fig:h0h1} the distribution of masses $(m_{H_1},m_{H_2})$ for the two lightest Higgs bosons. 
The strongest constraints in this plane come from the LHC upper limits on the reduced cross section of the Higgs decay $H\to\gamma\gamma$ \cite{cmsgamma,atlasgamma} and from the analogous LEP results for $H\to b\bar b$ \cite{Barate:2003sz}. The former puts a lower bound $m_{H_2}\gtrsim 122$ GeV on the mass of the SM-like Higgs, whereas the latter constrains vacua where the lightest Higgs $H_1$ couples too strongly to the SM fermions. Since in these models the scalar singlet component is distributed among $H_1$ and $H_2$, and large masses $m_{H_2}$ require a non-negligible singlet component for $H_2$, the cross section $\sigma^{b\bar b Z}(e^+e^-\to H_1Z)$ increases with the mass of $H_2$ and therefore LEP bounds effectively constrain $m_{H_2}$ from above, see figure \ref{fig:h0h1}. This upper bound on $m_{H_2}$ is in particular (slightly) stronger than the one derived from the recent LHC results. Vacua that are consistent with both LHC and LEP bounds thus have approximately  $m_{H_1}=100 \pm 15$ GeV and $m_{H_2}=124\pm 2$ GeV, with the signal of $H_1$ fitting in the $2\sigma$ excess observed at LEP \cite{Barate:2003sz}.

\begin{figure}[!t]
\begin{center}
\vspace{0.3cm}
\includegraphics[width=15.2cm]{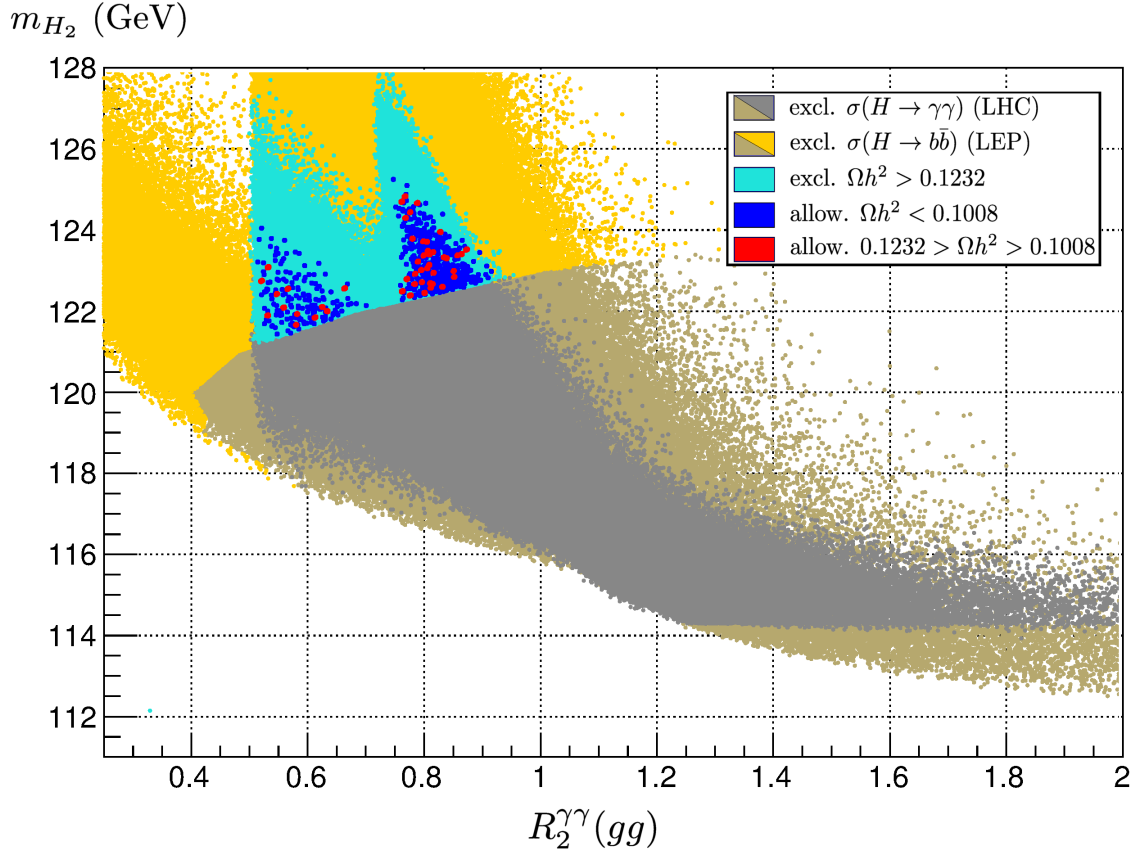}
\caption{\small Distribution of the the reduced signal cross section in the $H_2\to \gamma\gamma$ channel, $R_2^{\gamma\gamma}(gg\to H_2)$ versus the mass for the SM-like Higgs $H_2$ in NMSSM vacua with F-theory unification boundary conditions (\ref{fgut}) and unconstrained values of $A_\lambda$, $A_\kappa$ and $\lambda$. Dark blue and red points both pass all the current experimental and theoretical constraints. Dark blue points, however, have a deficiency of neutralino relic density and therefore require some additional source of dark matter.\label{fig:h1h1gg}}
\end{center}
\end{figure}

It is also interesting to compare the predicted reduced cross sections of the SM-like Higgs boson $H_2$ with the recently observed Higgs signal at the LHC, particularly in the $H\to \gamma\gamma$ channel, as it is starting to being measured with increasing precision. In figure \ref{fig:h1h1gg} we represent the reduced signal cross section in the $gg\to H_2\to \gamma\gamma$ channel
\begin{equation}
R_2^{\gamma\gamma}(gg)\equiv \frac{\sigma^{\gamma\gamma}(gg\to H_2)}{\sigma^{\gamma\gamma}_{\rm SM}(gg\to H)}
\end{equation}
against the mass of ${H_2}$. 
As we have already mentioned, the singlet component of $H_2$ is not negligible for the above allowed range of masses. This leads to a mild suppression of the couplings of the SM-like Higgs $H_2$ to the other SM fields, and in particular to the top quark and the $W$ bosons that dominate the one loop SM contribution to $\sigma^{\gamma\gamma}(gg\to H)$. Moreover, the stau is not light enough to enhance the di-photon production by running in the loops. Hence, as it can be observed in figure \ref{fig:h1h1gg}, there is no enhancement in $\sigma^{\gamma\gamma}(gg\to H_2)$ with respect to the SM but rather a mild suppression, with $R_2^{\gamma\gamma}(gg\to H_2)\simeq 0.7 - 0.9$. Although this is still in reasonable agreement with the latest LHC results \cite{cmsgamma,atlasgamma}, an experimental confirmation of a large enhancement in $R_2^{\gamma\gamma}(gg\to H_2)$ 
could disfavour the present class of models.\footnote{Nevertheless, as noted in appendix \ref{app}, mild enhancements in the di-photon rate with respect to the SM are allowed by keeping the computation of the Higgs masses at the one loop level. In this regard it would be interesting to understand the effect of the complete set of two loop corrections to the NMSSM Higgs sector on the di-photon rate.} For the same above reasons, similar considerations apply to the other reduced cross sections of $H_2$, leading also to mild  suppressions with respect to the SM. Concretely, we observe $R_2^{W^+W^-}(gg\to H_2)\sim R_2^{ZZ}(gg\to H_2)\sim R_2^{\tau^+\tau^-}({\rm VBF}\to H_2)\simeq 0.7 - 0.8$.

\begin{figure}[!t]
\begin{center}
\vspace{0.3cm}
\includegraphics[width=15.2cm]{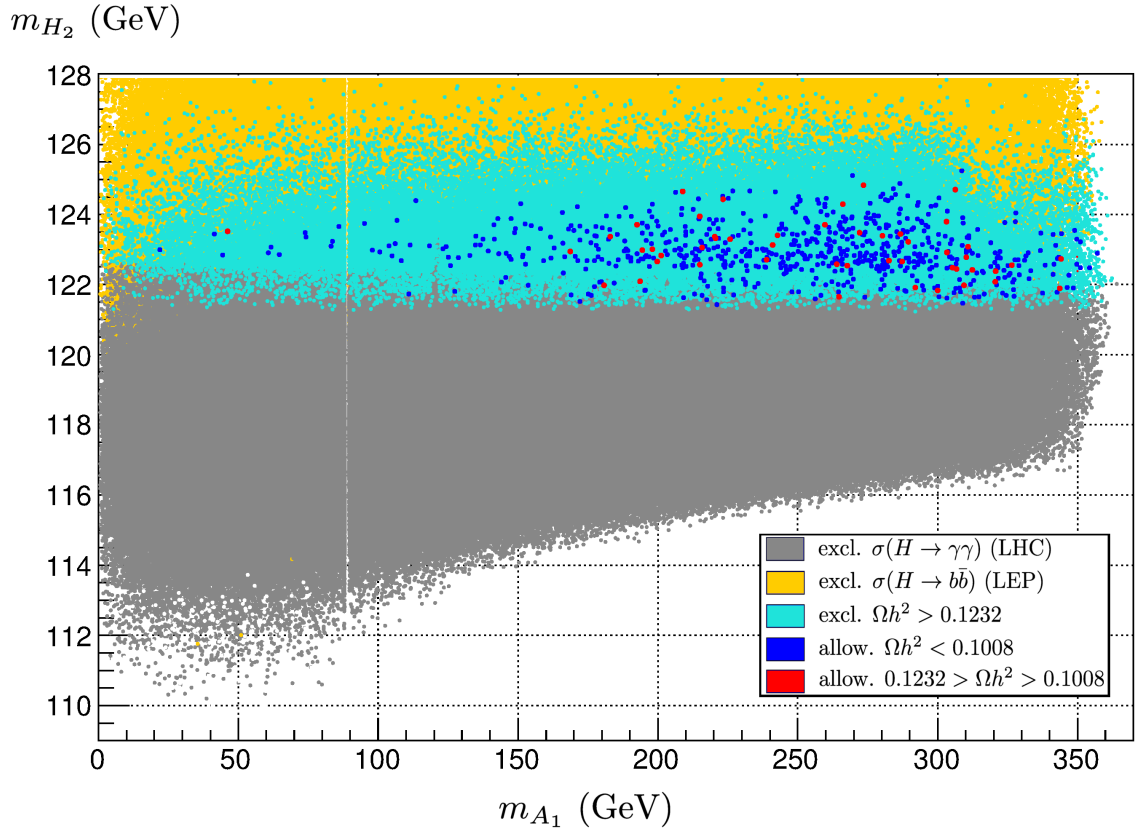}
\caption{\small Distribution of the SM-like Higgs and the lightest pseudoscalar masses $(m_{H_2},m_{A_1})$ for NMSSM vacua with F-theory unification boundary conditions (\ref{fgut}) and unconstrained values of $A_\lambda$, $A_\kappa$ and $\lambda$. Dark blue and red points both pass all the current experimental and theoretical constraints. Dark blue points, however, have a deficiency of neutralino relic density and therefore require some additional source of dark matter.\label{fig:h1ha}}
\end{center}
\end{figure}

We now briefly comment on the remaining part of the Higgs spectrum. Apart from $H_1$ and $H_2$ discussed above, the pseudoscalar $A_1$ is also relatively light in these models, with $m_{A_1}\lesssim 350$ GeV, see figure \ref{fig:h1ha}. However, it is highly dominated by its singlet component (its doublet composition is insignificant, of the order of $\simeq 10^{-8}\ \%$) and therefore  very decoupled from the SM fermions. In this limit the pseudoscalar mass can be written as $m_{A_1}^2\approx -3\kappa v_sA_\kappa$.
On the contrary, $H_3$, $A_2$ and $H^+$ have much larger masses and are nearly degenerated, with $m_{H_3}\simeq m_{A_2}\simeq m_{H^+}\gtrsim 1$~TeV. They are therefore much harder to detect at the LHC.

Finally, we may wonder about the robustness of the above Higgs spectrum. As we have already mentioned, in this region of the parameter space of the NMSSM there are relatively large uncertainties coming from some of the two loop corrections to the NMSSM Higgs sector that have not yet been computed. In this regard and for completeness, in appendix \ref{app} we perform the same analysis of this subsection but keeping only one loop radiative corrections to the Higgs sector into account. Such analysis reveals that the above qualitative results for the Higgs sector also hold at the one loop level, although the range of masses is broadened considerably and small enhancements to the di-photon rate with respect to the SM appear also to be possible.

\subsection{Neutralino dark matter}
\label{subsec:dark}

The lightest neutralino, $\neut$, can be a viable dark matter candidate in the NMSSM with interesting phenomenological properties \cite{Cerdeno:2004xw}. 
It differs from the MSSM in that it now contains a singlino component, which alters its couplings to the SM particles. 
The neutralino is also very sensitive to the structure of the Higgs sector, since it determines its annihilation cross section in the Early Universe (and thus the theoretical predictions for its relic density) and plays an important role in the computation of the scattering cross section of quarks.

In our construction, the neutralino turns out to be an almost pure singlino in all the allowed regions of the parameter space. 
This is due to the very small values of $\kappa$, which are at least one order of magnitude smaller than $\lambda$ as we already showed in figure~\ref{fig:lambdakappa}. This leads to a hierarchical structure in the neutralino mass matrix, $\kappa v_s\ll M_{1,2}<\lambda v_s$, which implies that the gaugino and Higgsino components of the lightest neutralino are almost negligible (to less than approximately a 0.1\%). 
This also implies that the neutralino mass can be small in these scenarios without violating any experimental limit. 
To a good approximation we can write $m_{\neut}\approx 2\kappa v_s$, and in principle, neutralinos as light as 50~GeV are possible, but when the recent experimental constraints are applied on other observables (mostly the Higgs decays discussed above), we are left with the range $100\textrm{ GeV}\lsim m_{\neut}\lsim 240$~GeV. 

\begin{figure}[!t]
\begin{center}
\vspace{0.3cm}
\includegraphics[width=15.2cm]{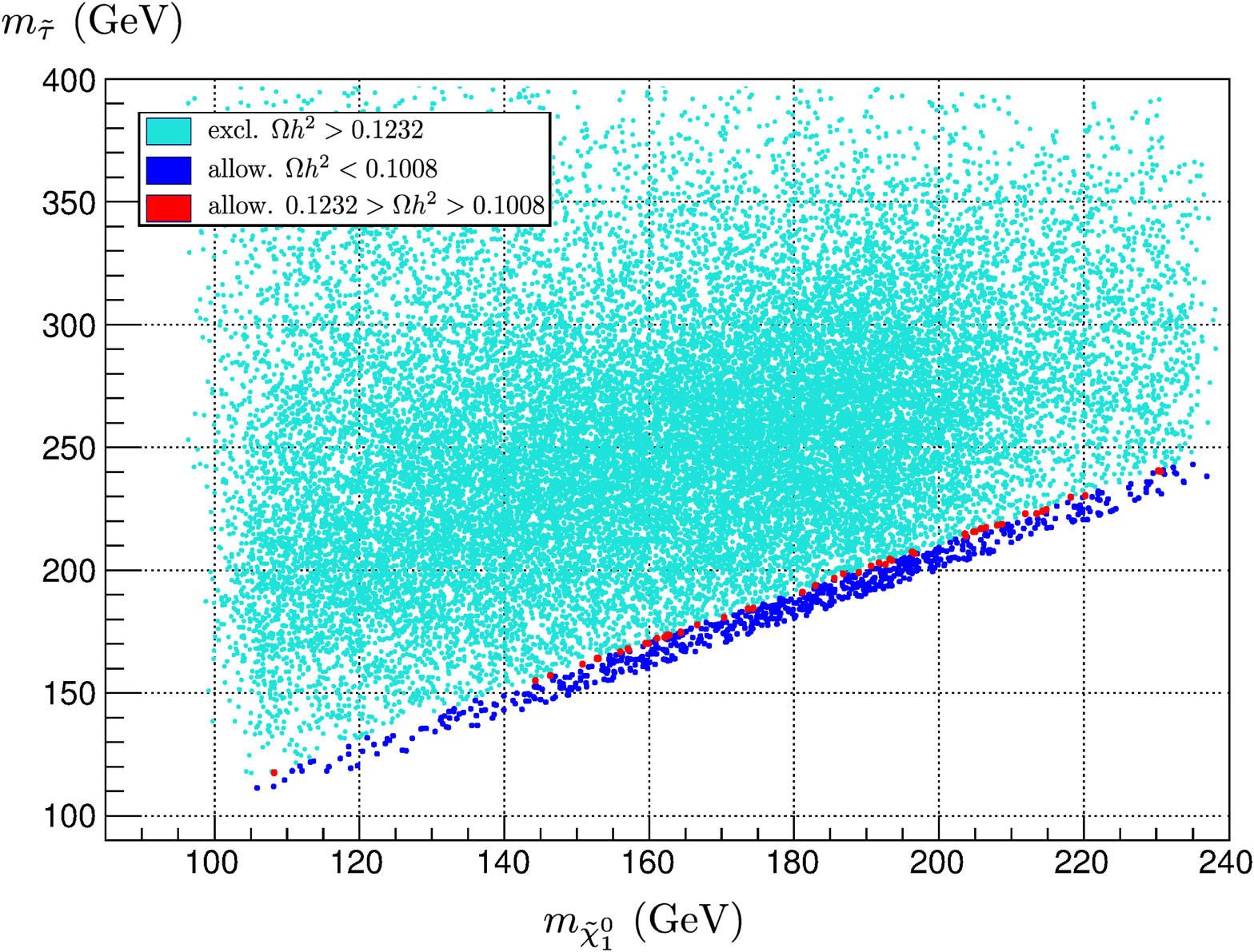}
\caption{\small Distribution of neutralino and stau masses $(m_{\neut},m_{\tilde \tau})$ for NMSSM vacua with F-theory unification boundary conditions (\ref{fgut}) and unconstrained values of $A_\lambda$, $A_\kappa$ and $\lambda$, for points that pass  all current experimental constraints except that of the relic density, that is encode in the legend of colors.\label{fig:stauchi}}
\end{center}
\end{figure}

We have represented in figure~\ref{fig:stauchi} the mass of the lightest neutralino $m_{\neut}$ against the stau mass $m_{\tilde \tau}$, for vacua that pass all current experimental constraints except that of the relic density observed by WMAP. A singlino-like neutralino has reduced couplings to the SM particles and thus generally displays a small annihilation cross-section in the Early Universe. 
As a consequence, the relic density is too large in most of the parameter space, and typically exceeds the WMAP range $ 0.1008 < \Omega h^2 < 0.1232 $ (at the $2\sigma$ level). 
The relic density can be lowered through coannihilation effects when the stau mass is very close to that of the neutralino. In fact, we observe that the allowed regions of the parameter space have $m_{\tilde \tau}-m_{\neut}\approx10$~GeV. This leads to a very interesting prediction of this scenario, namely that the stau mass is in the range $110 \textrm{ GeV}\lsim m_{\tilde\tau}\lsim 250$~GeV. 
Notice that unlike the scenarios discussed in ref.\,\cite{Aparicio:2012iw}, the stau-neutralino mass difference is never small enough to allow for long-lived staus.

Dark matter can be detected directly through its scattering off nuclei in a detector. 
The interaction of neutralinos with quarks has contributions from $s$-channel squark exchange and $t$-channel Higgs exchange diagrams. 
Since the neutralino is a pure singlino state, its coupling to squarks is completely negligible. A pure singlino only couples to the singlet part of the Higgs, but this coupling $C_{\tilde S S S}=-2\sqrt{2}\kappa$ vanishes in the limit $\kappa\to 0$ and is tiny in our vacua.
This results in a extremely small neutralino-nucleon scattering cross-section. In particular, the theoretical predictions for the spin-independent contribution is below $\sigma_{\neut-p}^{SI}\approx10^{-13}$~pb, several orders of magnitude below the predicted sensitivity of the projected 1 ton scale detectors. The same happens with the spin-dependent component.

\subsection{Constraints on rare decays and the muon anomalous magnetic moment}
\label{subsec:bmumu}

Rare decays constitute excellent probes for new physics beyond the SM. In particular, the effect of flavour-changing neutral currents in $b$-physics signals is extremely interesting in supersymmetric models, since it can be sizeable at large $\tan\beta$, and rather sensitive to the Higgs sector.
For example, supersymmetric contributions to the branching ratios of the rare processes $\bmumu$ or $\bsg$ can easily exceed the experimental measurements of these quantities and generally lead to stringent constraints on the parameter space. 
This is particularly important in the NMSSM due to the presence of new scalar and pseudoscalar Higgsses which induce new contributions to these observables.

Let us start by addressing the branching ratio BR($\bmumu$). 
This observable can be written in terms of the Wilson coefficients which appear the effective Hamiltonian that describes the transition $b\to s$ as follows, 
\begin{equation}
{\rm BR}(\bmumu) \sim \left[\left(1-\frac{4 m_\mu^2}{m_B^2}\right)\,C_S^2 + \left(C_P + \frac{2 m_\mu}{m_B^2}\,C_A\right)^2\right]\ .
\end{equation}
In the SM calculation of this quantity \cite{Buras:2002vd} only $C_A$ is relevant, since $C_S$ and $C_P$ are suppressed by the small Yukawas.  
However, in supersymmetric theories, there are penguin contributions involving the neutral scalar and pseudoscalar Higgs bosons to $C_S$ and $C_P$ which can be sizeable, both in the MSSM \cite{Bobeth:2001jm} and the NMSSM  \cite{Hiller:2004ii,Domingo:2007dx,Hodgkinson:2008qk}. 
In the MSSM one finds BR$(\bmumu)\propto(\tan^6\beta/m_A^4)$, and therefore in the large $\tan\beta$ regime supersymmetric contributions easily exceed the recent LHCb measurement BR($\bmumu$)$=3.2{+1.5 \atop -1.2}\times 10^{-9}$ \cite{:2012ct}, especially for light pseudoscalars. In the NMSSM the Wilson coefficient $C_P$ receives contributions from both pseudoscalar Higgses, but it is only their doublet component that contributes.

As we have already commented in section \ref{subsec:higgs}, the viable points resulting from our scan display a relatively light pseudoscalar, $m_{A_1}\lesssim 350$~GeV, which is a pure singlet. Therefore, it does not contribute to the Wilson coefficient $C_P$, and only the heavy pseudoscalar $A_2$ has to be taken into account. The latter is rather heavy (in our scan its mass varies in the range from $1$ to $1.4$~TeV) and as a consequence the resulting value of $C_P$ is also small.
In fact, it is of the same order as the SM contribution $(2 m_\mu/m_B^2)\,C_A$ but with opposite sign. This implies that not only BR$(\bmumu)$ can be small, in agreement with the experimental data, but even a cancellation between $C_P$ and $C_A$ that leads to a smaller value than the SM prediction is possible  (see \cite{Elor:2012ig} for a similar effect in the context of the MSSM). 
This last possibility, in fact, happens in parts of our parameter space for $M\gsim 1200$~GeV, where values as low as BR$(\bmumu)\approx 1.8\times10^{-9}$ can be obtained. As we will see in the next section (see figure \ref{fig:m12tanb}), for not too large values of tan\,$\beta$ this actually favours a light spectrum.

We will now briefly turn our attention to the branching ratio BR($\bsg$). In the past years this has been one of the strongest constraints on supersymmetric models, mainly because of the additional contributions from loops of charged Higgs bosons. There are also specific contributions from the extended Higgs and neutralino sectors of the NMSSM, although these start at the two loop level \cite{Hiller:2004ii,Domingo:2007dx}. Whereas imposing the experimental result BR$(\bsg)= (3.52 \pm 0.23 \pm 0.09) \times 10^{-4}$ generally leads to constraints on the NMSSM parameter space \cite{Cerdeno:2007sn}, our scan reveals that these bounds are now superseded by those on BR$(\bmumu)$ discussed above. 
Therefore, the current experimental bounds on BR$(\bsg)$ do not have a large impact in the space of parameters of these models.

Let us finally address the supersymmetric contribution to the muon anomalous magnetic moment, $a_\mu^{\rm SUSY}$.  
The observed discrepancy between the experimental value \cite{g-2} and the Standard Model predictions using $e^+e^-$ data, favours positive contributions from new physics. These can be constrained to be in the range $10.1\times10^{-10}<a_\mu^{SUSY}<42.1\times10^{-10}$ at the $2\,\sigma$ confidence level \cite{Hagiwara:2011af} where theoretical and experimental errors are combined in quadrature (see also refs.~\cite{Jegerlehner:2009ry,Davier:2010nc} that lead to a similar range). 
However, if tau data is used this discrepancy is smaller $2.9\times10^{-10}<a_\mu^{SUSY}<36.1\times10^{-10}$ \cite{Davier:2010nc}.
The theoretical predictions for this observable generally increase for light supersymmetric masses and large $\tan\beta$. 
Although in this scenario the masses of the first and second family sleptons are relatively high, we are in the regime of large $\tan\beta$. The predicted $a_\mu^{\rm SUSY}$ turns out to be within the $3\,\sigma$ range of the $e^+e^-$ data and well within the $2\,\sigma$ range of the discrepancy predicted by tau data.

To illustrate this discussion, we display the specific values of these low energy observables for our choice of representative benchmark points in table \ref{tab:observables}.

\begin{table}[!t] \footnotesize
\renewcommand{\arraystretch}{1.50}
\begin{center}
\tabcolsep 3.8pt
\begin{tabular}{|c||c|c|c|c|c|c|c|}
\hline 
Point & BR($B_s \rightarrow \mu^+ \mu^-$) & BR($b \rightarrow s \gamma$) & $a_\mu^{\rm SUSY}$ & $R^{\gamma\gamma}_{H_1}(gg)$ & $R^{\gamma\gamma}_{H_2}(gg)$ & $R^{VV}_{H_1}(gg)$ & $R^{VV}_{H_2}(gg)$     \\
\hline\hline
 $P_{1}$ & $3.17 \cdot 10^{-9}$ & $2.88 \cdot 10^{-4}$ & $8.32\cdot 10^{-10}$ & 0.250 & 0.775 & 0.242 & 0.742   \\
 \hline
 $P_{2}$ & $3.40 \cdot 10^{-9}$ & $2.89 \cdot 10^{-4}$ & $8.05 \cdot 10^{-10}$ & 0.263 & 0.798 & 0.243 & 0.739    \\
 \hline
 $P_{3}$ & $2.62 \cdot 10^{-9}$ & $2.95 \cdot 10^{-4}$ & $6.66 \cdot 10^{-10}$ & 0.280 & 0.766 & 0.262 & 0.722 \\
 \hline
$P_{4}$ & $3.77 \cdot 10^{-9}$ & $2.97 \cdot 10^{-4}$ & $6.73 \cdot 10^{-10}$ & 0.288 & 0.767 & 0.267 & 0.715   \\
 \hline
$P_{5}$ & $2.86 \cdot 10^{-9}$ & $2.98 \cdot 10^{-4}$ & $6.22 \cdot 10^{-10}$ & 0.280 & 0.761 & 0.264 & 0.721   \\
 \hline
$P_{6}$ & $2.38 \cdot 10^{-9}$ & $3.03 \cdot 10^{-4}$ & $5.00 \cdot 10^{-10}$ & 0.302 & 0.774 & 0.275 & 0.708  \\
 \hline
$P_{7}$ & $3.31 \cdot 10^{-9}$ & $3.04 \cdot 10^{-4}$ & $4.99 \cdot 10^{-10}$ & 0.248 & 0.813 & 0.228 & 0.754 \\
 \hline
$P_{8}$ & $2.35 \cdot 10^{-9}$ & $3.05 \cdot 10^{-4}$ & $4.48 \cdot 10^{-10}$ & 0.297 & 0.804 & 0.264 & 0.718  \\
 \hline
\end{tabular}
\end{center} \caption{Low energy observables \label{tab:observables} and reduced cross-sections for different Higgs signals calculated for the set of benchmark points.}
\end{table}

\section{Consistency with F-theory unification}
\label{sec:consftheory}

In the previous section we have discussed the main phenomenological constraints that shape the parameter space of NMSSM vacua with F-theory unification boundary conditions eqs.~(\ref{fgut}). We now describe the overall structure of the allowed regions and further comment on their consistency with an underlying F-theory unification structure, including also the extra conditions (\ref{singletgut}). 

Indeed, as we have already mentioned, leaving the singlet sector unconstrained, NMSSM vacua with F-theory unification boundary conditions are completely specified by six parameters that encode the relevant information of the local F-theory background. In our scan these are given by the values of $M$, $\rho_H$, $A_\lambda$ and $A_\kappa$ at the unification scale, $\lambda$ at the supersymmetry breaking scale and $\textrm{tan }\beta$ at the scale $m_Z$.  
However, in specific F-theory GUTs the singlet sector will also satisfy a set of boundary conditions at the unification scale which, under the simplifying assumptions of section \ref{sec:ftheory}, are given by eqs.~(\ref{singletgut}). Hence, the number of independent parameters in specific models can be reduced from six to three (for instance, the values of $M$, $\rho_H$ and $\lambda$ at the unification scale), as we discuss below.

\subsection{$M$, $\rho_H$ and $\textrm{tan }\beta$}
\label{sec:mrho}

\begin{figure}[!t]
\begin{center}
\vspace{0.3cm}
\includegraphics[width=15cm]{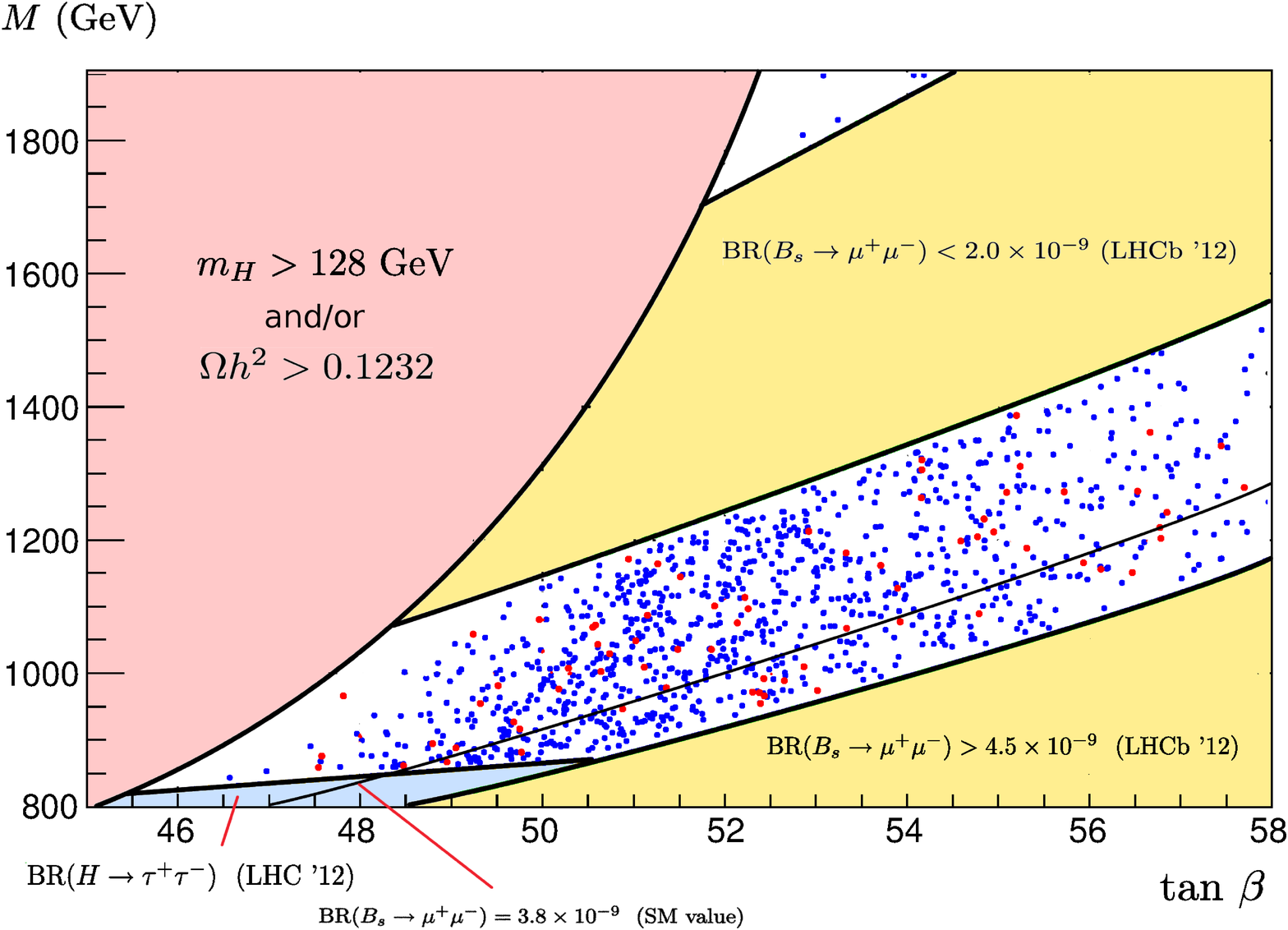}
\caption{\small Distribution of NMSSM vacua over the plane $(M,\, \textrm{tan }\beta)$ with F-theory boundary conditions at the unification scale given by eqs.~(\ref{fgut}) and unconstrained values of $A_\lambda$, $A_\kappa$ and $\lambda$.  Points pass all the current experimental and theoretical constraints. Dark blue points, however, have a deficiency of neutralino relic density and therefore require some additional source of dark matter.\label{fig:m12tanb}}
\end{center}
\end{figure}

Let us first address the effect of the phenomenological constraints on the region of the parameter space spanned by $M$, $\rho_H$ and $\textrm{tan }\beta$. We have represented in figure \ref{fig:m12tanb} the distribution of phenomenologically viable NMSSM vacua with F-theory unification boundary conditions (\ref{fgut}), over the relevant region of the plane $(M,\, \textrm{tan }\beta)$. 
As we have advanced in previous sections, vacua satisfying all the current phenomenological constraints sit in the region of large $\textrm{tan }\beta$, with $\textrm{tan }\beta \gtrsim 46$.  In this region the contribution of the heaviest pseudoscalar $A_2$ to the branching ratio $BR(B_s\to \mu^+\mu^-)$ is such that it leads to important cancellations with the SM contribution, giving rise to comparable or even reduced values with respect to the SM (see section \ref{subsec:bmumu}). For regions of the parameter space where the above cancellation is maximal the  branching ratio $BR(B_s\to \mu^+\mu^-)$  in particular becomes too small, below the recent 95 \% confidence level lower limit measured by the LHCb collaboration \cite{:2012ct}. For moderately large values of tan\,$\beta$, this disfavours a band in the plane $(M,\, \textrm{tan }\beta)$ with approximately $1200 \textrm{ GeV} \lesssim M \lesssim 1800$~GeV, see figure \ref{fig:m12tanb}. Nevertheless, the reader should bear in mind that all points in this band have $BR(B_s\to \mu^+\mu^-)\gtrsim 1.8 \times 10^{-9}$, and therefore many of them are actually within the theoretical uncertainties from QCD \cite{Bobeth:2001jm}.

Regarding the scale of supersymmetry breaking, we observe in the same figure that the lower bound on $M$ compatible with all the current experimental constraints is $M\simeq 850\ \textrm{GeV}$. Below that scale, the cross section for the SM-like Higgs decay $H_2\to\tau^+\tau^-$ becomes larger than the most recent experimental upper limits set by the CMS and ATLAS collaborations \cite{htautau,htautauatlas}. On the other hand, since the LSP is a singlino-like neutralino, its mass $m_{\tilde \chi^0_1}\simeq 2\kappa s\lesssim 240 \ \textrm{GeV}$ is independent of $M$ and vacua with the correct neutralino relic density are possible for arbitrarily large values of $M$, contrary to what occurs in the modulus dominated MSSM \cite{Aparicio:2008wh,Aparicio:2012iw}. Nevertheless, the amount of fine-tuning required in order to have a light stau with mass in the region of co-annihilation increases considerably for large values of $M$.\footnote{Moreover, for large values of $M$ the value of $\textrm{tan }\beta$ becomes too large for the computation of down-type Yukawa couplings to be reliable.} 

\begin{figure}[!t]
\begin{center}
\vspace{0.3cm}
\includegraphics[width=15cm]{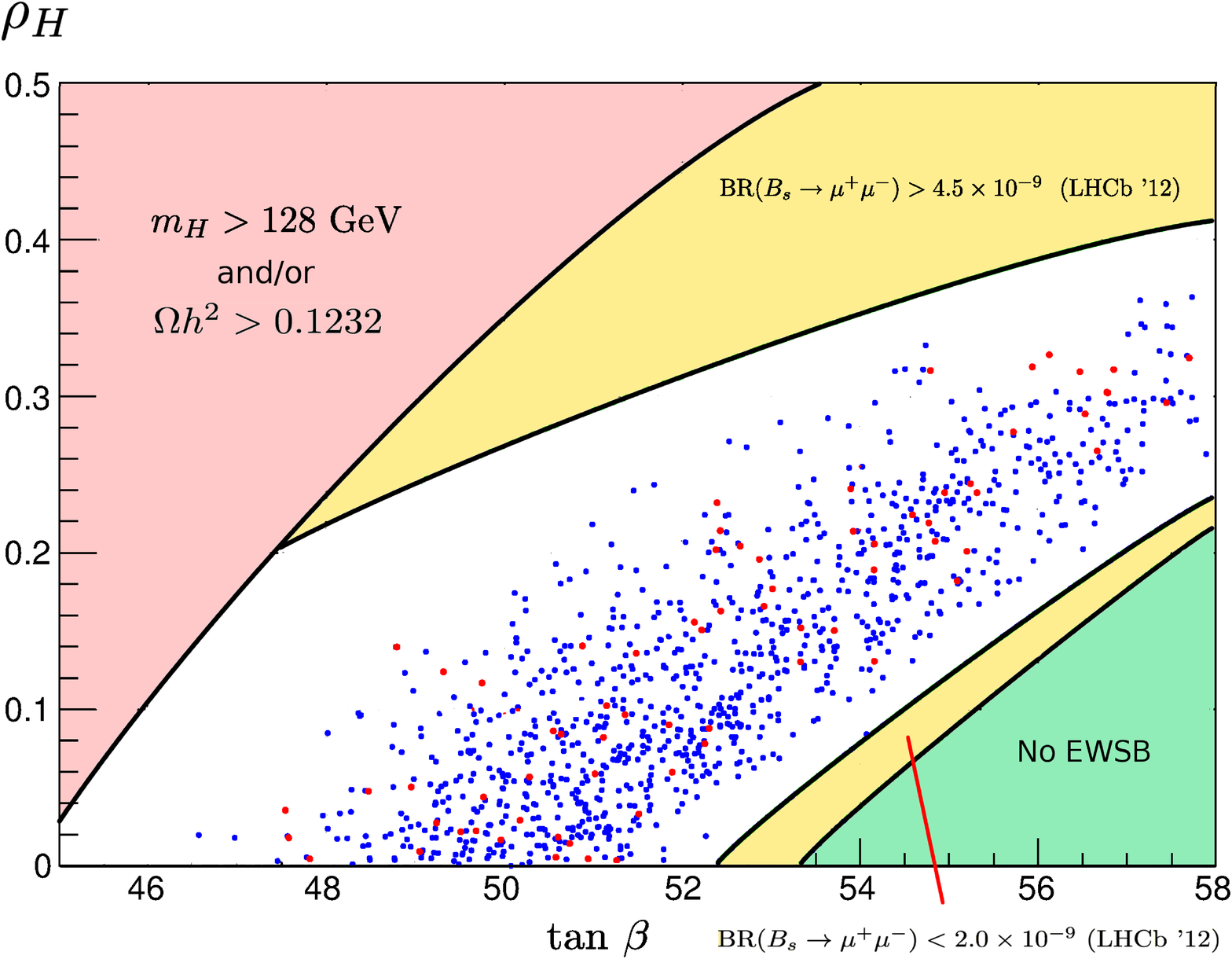}
\caption{\small Distribution of NMSSM vacua over the plane $(\rho_H,\,\textrm{tan }\beta)$ with F-theory boundary conditions at the unification scale given by eqs.~(\ref{fgut}) and unconstrained values of $A_\lambda$, $A_\kappa$ and $\lambda$. Points pass all the current experimental and theoretical constraints. Dark blue points, however, have a deficiency of neutralino relic density and therefore require some additional source of dark matter. The upper yellow and pink regions in this plot are only indicative: whereas \emph{all} points inside this regions are excluded because the branching ratio BR($B_s\to \mu^+ \mu^-$), $m_{H_2}$ and/or $\Omega h^2$, there are also points outside this regions that are also excluded by these observables. \label{fig:rhotanb}}
\end{center}
\end{figure}

Similarly, we have represented  in figure \ref{fig:rhotanb}  the distribution of phenomenologically viable vacua on  the plane $(\rho_H,\, \textrm{tan }\beta)$. 
In this case, the LHCb upper limit on $BR(B_s\to \mu^+\mu^-)$ forces $\rho_H$ to take small values, $\rho_H \lesssim 0.4$,
with most of the points below 0.2. Remarkably, this is consistent with our treatment of the effect of magnetic fluxes on the K\"ahler metrics for the Higgs as small perturbations, see section \ref{sec:ftheory}.

Having looked at the effect of the phenomenological constraints on $M$, $\rho_H$ and $\textrm{tan }\beta$, let us now move on to the discussion of the singlet sector.

\subsection{The singlet sector: $A_\lambda$, $A_\kappa$ and $\lambda$}
\label{subsec:singlet}

As already mentioned in section \ref{sec:ftheory}, the singlet $S$ is not localized in the local 4-cycle $\mathcal{S}$ and therefore its interactions depend on the details of the background along the bulk of the compactification, making this sector very model-dependent. Nevertheless, in simple models where the only source of supersymmetry breaking for $S$ is given by the F-terms $F_{t}$ and $F_{t_b}$ we expect $A_\lambda$, $A_\kappa$ and $m_S$ to satisfy a set of boundary conditions at the unification scale which, to first approximation, are given by eqs.~(\ref{singletgut}). 

\begin{figure}[!t]
\begin{center}
\vspace{0.3cm}
\includegraphics[width=15cm]{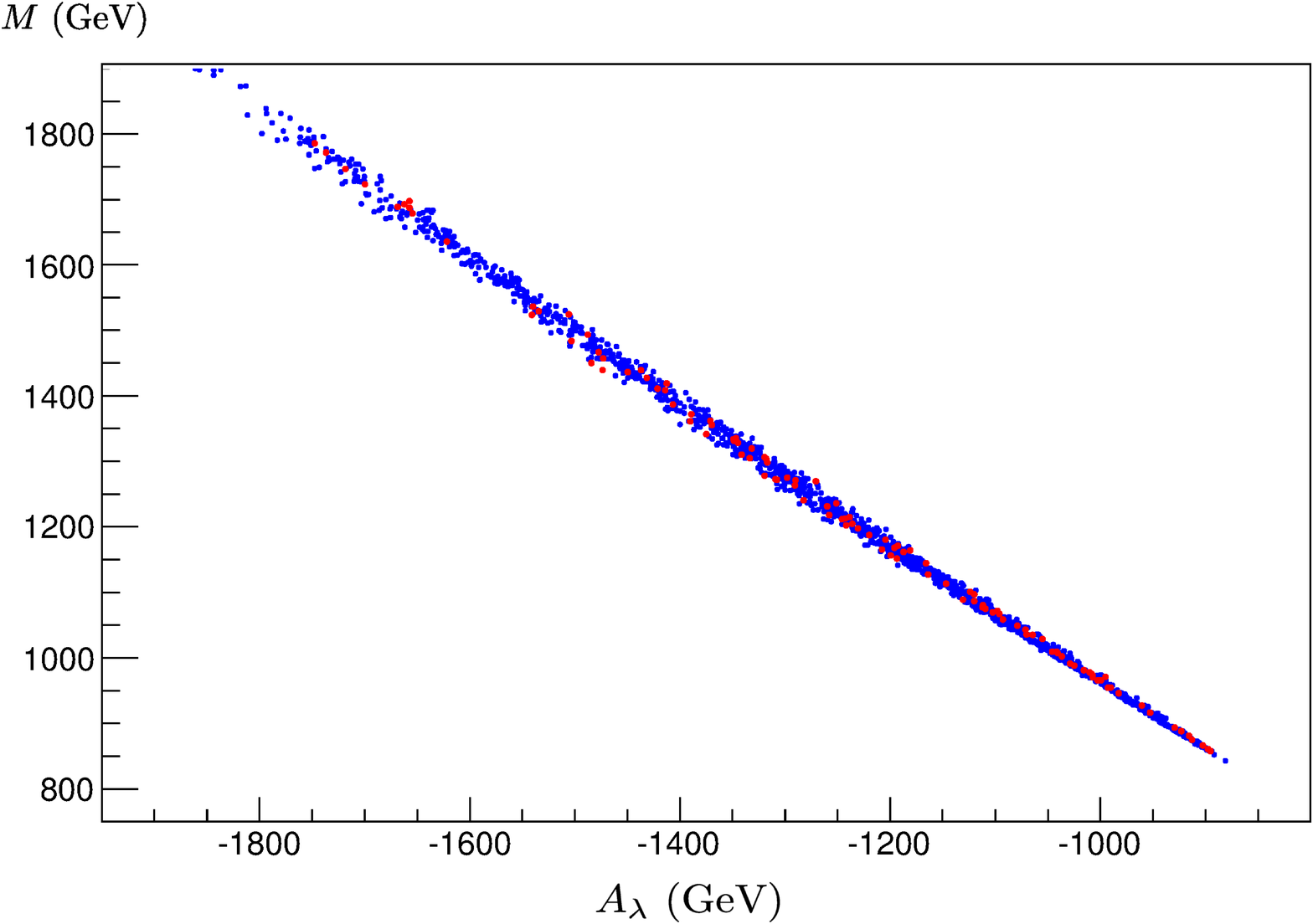}
\caption{\small Distribution of NMSSM vacua over the plane $(M,\, A_\lambda)$ with F-theory boundary conditions at the unification scale given by eqs.~(\ref{fgut}) and unconstrained values of $A_\lambda$, $A_\kappa$ and $\lambda$. Points pass all the current experimental and theoretical constraints. Dark blue points, however, have a deficiency of neutralino relic density and therefore require some additional source of dark matter. For convenience, in this plot we have not imposed the lower bound BR$(\bmumu)>2.0\times10^{-9}$. \label{fig:linea}}
\end{center}
\end{figure}

It is highly remarkable to see how the above type of relations are also strongly favoured by the current experimental input. Indeed, although we have scanned over unconstrained values and sign choices of $A_\lambda$, $A_\kappa$ and $\lambda$, requiring a SM-like Higgs in the window $122 - 128$ GeV and a neutralino relic density below the WMAP bound $\Omega h^2<0.1232$ uniquely selects the sign choice $A_\lambda<0$, $A_\kappa\leq 0$ and $\lambda\geq 0$, consistently with the F-theoretic boundary conditions for the singlet sector, eqs.~(\ref{singletgut}). Moreover, it is possible to check that the current experimental constraints imply also a strong correlation between the values of $A_\lambda$ and $M$ at the unification scale, similar to that of eqs.~(\ref{singletgut}). Indeed, in figure \ref{fig:linea} we have represented the distribution of NMSSM vacua on the plane $(M,\, A_\lambda)$ with  boundary conditions (\ref{fgut}) and unconstrained values of $A_\lambda$, $A_\kappa$ and $\lambda$. The linear correlation between $A_\lambda$ and $M$ for vacua that pass all the current phenomenological constraints is clearly visible in this figure. 
Finally, in all these vacua $A_\kappa$ and $m_S$ take relatively small values at the unification scale, with $-0.015 \lesssim (m_S/m_{\bf{\bar 5},{\bf 10}})^2  \lesssim 0.035$ and $(A_\kappa/A) \lesssim 0.3$ (see table~\ref{table:gut} in section~\ref{sec:signatures}), also in fairly good agreement with the F-theoretic expectation $A_\kappa\simeq m_S\simeq 0$ in eqs.~(\ref{singletgut}).

\begin{figure}[!t]
\begin{center}
\vspace{0.3cm}
\includegraphics[width=15.3cm]{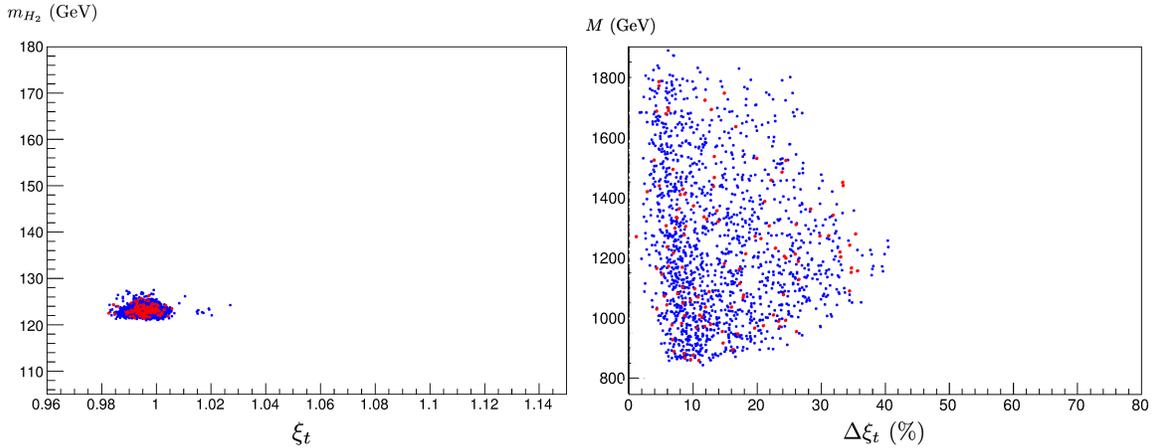}
\caption{\small (Left): Modular weight $\xi_t$ estimated from eqs.~(\ref{singletgut2}) against the mass of the SM-like Higgs, for NMSSM vacua with F-theory boundary conditions at the unification scale given by eqs.~(\ref{fgut}) and unconstrained values of $A_\lambda$, $A_\kappa$ and $\lambda$. (Right): Percentage of mismatch between the three eqs.~(\ref{singletgut2}) against $M$. For convenience, in this plot we have not imposed the lower bound BR$(\bmumu)>2.0\times10^{-9}$. 
In both figures all the points pass all the current experimental and theoretical constraints. Dark blue points, however, have a deficiency of neutralino relic density and therefore require some additional source of dark matter.\label{fig:singletfit}}
\end{center}
\end{figure}

In order to make a more quantitative estimation of the compatibility between  the phenomenological constraints and the F-theoretic relations for the singlet sector, we can consider a slightly more general K\"ahler metric for $S$ than that of eq.~(\ref{metricascurvas}), depending also on the K\"ahler modulus of $\mathcal{S}$,
\begin{equation}
K_S=\frac{t_S^{1/2}\, t^{1-\xi_t}}{t_b}
\end{equation}
Making use of this metric, the boundary conditions (\ref{singletgut}) generalize to
\begin{align}
A_\lambda&=-M(2-\xi_t-\rho_H)\label{singletgut2}\\
A_\kappa&= -3M(1-\xi_t)\nonumber\\
m_S^2&=|M|^2(1-\xi_t)\nonumber
\end{align}
where the simple F-theory unification conditions for the singlet sector, eqs.~(\ref{singletgut}), are recovered for $\xi_t=1$. Making use of these equations we can estimate the modular weight $\xi_t$ in three different ways (one from each equation) for each vacuum in our scan with  boundary conditions (\ref{fgut}) and unconstrained values of $A_\lambda$, $A_\kappa$ and $\lambda$. If the vacuum is consistent with the F-theoretic boundary conditions, the three estimations of $\xi_t$ should agree with each other and with $\xi_t\simeq 1$. We have represented in figure \ref{fig:singletfit} (left) the maximum value of the three estimations against the mass of the SM-like Higgs $H_2$ for vacua that pass all the current experimental constraints. We observe in this figure that the latter strongly favour the F-theory value $\xi_t=1$. 
Similarly, in figure \ref{fig:singletfit} (right) we have represented the maximum percentage of mismatch between the three different estimations of $\xi_t$, showing that for a large fraction of the phenomenologically viable vacua the mismatch between the three eqs.~(\ref{singletgut2}) is lower than 10\%. 

Hence, whereas in the scan over modulus dominated vacua that we have performed in section \ref{subsec:scan} we have taken six independent parameters, it turns out that many of the vacua that satisfy all the current experimental bounds can be actually described to a large extend in terms of just three parameters, namely $M$, $\rho_H$ and $\lambda$, satisfying the boundary conditions (\ref{fgut}) together with the additional boundary conditions (\ref{singletgut}) for the singlet sector.

\section{Supersymmetric spectrum and signatures}
\label{sec:signatures}

The model is extremely predictive.
After imposing all the experimental constraints on the parameter space we are left with a very characteristic supersymmetric spectrum. In table\,\ref{tab:spectrum} we indicate the masses of the various supersymmetric particles and Higgs bosons for a series of representative benchmark points.

Regarding the Higgs sector, 
this model provides a very interesting scenario, with two light scalar Higgses. In this sense it seems similar to the proposal of ref.\,\cite{Belanger:2012tt} where the lightest Higgses had masses of 98 and 125~GeV. 
Notice however that the solution that we find belongs to a completely different region of the NMSSM parameter space. 
The most important differences are the large value of $\tan\beta$ in our scenario and the very small values of both $\lambda$ and $\kappa$. Also $\mu_{eff}\gsim 1$~TeV, thus being much larger than in ref.\,\cite{Belanger:2012tt}.
This leads to a different phenomenology (as already emphasized in section\,\ref{subsec:higgs}, most notably, in our case we predict no enhancement of Higgs decay $H_2\to\gamma\gamma$).
Notice also that this scenario does not contain exotic channels for Higgs decay of the two lightest CP-even states, since the pseudoscalar is heavier than $H_1$ and $H_2$. In any case, we can be certain that this scenario can be tested through the branching fractions of these two lightest states. In particular, the lightest Higgs could lead to a peak in the $H_1\to\gamma\gamma$ channel around 100 GeV that would be observable at the LHC. Regarding the lightest pesudoscalar, it is singlet like and its production rate is extremely suppressed. The heavier pseudoscalar and the third CP-even Higgs state are doublet like, but very heavy and thus very difficult to produce.

The coloured section is in the 1.5 to 3 TeV range.
Squarks are lighter than the gluino and, interestingly, the lightest stop can be as light as $m_{\tilde t_1}\approx 1200$~GeV in the regions with smaller gaugino masses. 
As already emphasized the next-to-lightest supersymmetric particle is the lightest stau, whose mass difference with the neutralino is of the order of 10~GeV for the whole range of viable gaugino masses. The nature of the neutralino and chargino sector is easily understood from the resulting hierarchy $\kappa v_s\ll M_{1}<M_2<\lambda v_s$. As already stated in section\,\ref{subsec:dark}, the lightest neutralino is pure singlino. The second neutralino state is bino-like and relatively light ($m_{\tilde\chi_2^0}\approx300-500$~GeV). The third neutralino state and the lightest chargino are wino-like and with a mass in the range $m_{\tilde\chi_3^0,\tilde\chi_1^\pm}\approx600-1000$~GeV. Finally the heavier neutralino states and the heavier chargino are Higgsino-like states with masses approximately equal to $\mu_{eff}=\lambda v_s$.

This kind of spectrum is very similar to what is obtained in the Constrained NMSSM \cite{Djouadi:2008uj} and both share the same search strategies. 
In particular, the branching ratios of the lightest chargino and the second and third neutralino states into the stau NLSP are sizable, thus potentially leading to significant rates of tau-rich final states \cite{Ellwanger:2010es,Belanger:2012jn}. 
Let us be more specific about this. 
Left-handed squarks can decay into the lightest chargino with a branching fraction of approximately 65\% and into the third neutralino with 33\%, whereas right-handed squarks decay mostly into the second neutralino with a branching fraction of approximately 99\%. Regarding the stops, the lightest stop undergoes the following decays $\tilde t_1\to \tilde\chi_2^0 t$ (60\%), $\tilde t_1\to \tilde\chi_3^0 t$ (12\%), $\tilde t_1\to \tilde\chi_1^\pm t$ (25\%). So either if they are directly produced of obtained in gluino decays, the resulting production of $\tilde\chi_1^{\pm}$ and $\tilde\chi_{2,3}^0$ is copious. 
The second and third neutralino states decay in turn into the lightest stau with a branching ratio of approximately 100\%. Obviously the stau NLSP can only decay into the lightest neutralino, producing another tau in the final state which is softer than the previous one (as the mass-difference of the stau and neutralino is approximately 10~GeV) thus leading to $\tilde\chi_{2,3}^0 \to\tilde\tau_1\tau\to\tau^+\tau^-\tilde\chi_1^0$. 
The lightest chargino decays into a stau and the corresponding neutrino thus giving rise to only one final state tau $\tilde\chi_1^{\pm}\to\nu_\tau\tilde\tau_1\to\nu_\tau\tau\tilde\chi_1^0$.
The signal expected from this kind of scenarios is therefore the presence of multitau signals, originated from the two chains of cascade decays, associated to the emission of hard central jets and missing energy \cite{Ellwanger:2010es}.

Notice finally that the upper bound set by BR($\bmumu$) implies that the whole spectrum is lighter than approximately $3$~TeV. This is well within the reach of LHC at 14~TeV for searches involving multijets plus missing energy.

\begin{table}[!t] \footnotesize
\renewcommand{\arraystretch}{1.50}
\begin{center}
\tabcolsep 3.8pt
\begin{tabular}{|c||c|c|c|c|c|c|c|c|c|c|c|}
\hline 
Point & ${\tilde g}$ & ${\tilde Q}_{R,L}$ & ${\tilde t}_{1,2}$ & ${\tilde b}_{1,2}$&$\tilde{L}_{R,L}$ & $\tilde\tau_{1,2}$& $\tilde\chi_i^0$ & $\tilde\chi_i^+$& $m_{H_i}$ & $m_{A_i}$ & $m_{H^+} $ \\
\hline\hline
 $P_{1}$ & 1921 & \caja{1758\\ 1827} & \caja{1263\\ 1558 }& \caja{1481\\ 1583}& \caja{684\\ 827} &\caja{230\\ 719}& \caja{213\\ 367; 696\\ 1238; 1243}& \caja{696\\1244} & \caja{103\\122.4\\ 1016} & \caja{321\\1016} & 1019 \\ \hline
 $P_{2}$ & 1983 & \caja{1814\\ 1886} & \caja{1302\\ 1601 }& \caja{1521\\ 1626}& \caja{708\\ 855} &\caja{175\\ 735}& \caja{164\\ 381; 721\\ 1274; 1279}& \caja{721\\1279} & \caja{98.1\\123.4\\ 1036} & \caja{220\\1036} & 1040 \\ \hline
 $P_{3}$ & 2716 & \caja{1989\\ 2069} & \caja{1434\\ 1749 }& \caja{1673\\ 1778}& \caja{782\\ 944} &\caja{199\\ 807}& \caja{189\\ 423; 800\\ 1394; 1400}& \caja{800\\ 1400} & \caja{96.9\\124.8\\ 1131} & \caja{273\\ 1131} & 1134 \\ \hline 
 $P_{4}$ & 2236 & \caja{2042\\ 2125} & \caja{1499\\ 1802}& \caja{1718\\ 1827}& \caja{804\\ 971} &\caja{197\\ 831}& \caja{186\\ 436; 824\\ 1440; 1444}& \caja{824\\ 1445} & \caja{97.4\\124.3\\ 1095} & \caja{266\\ 1094} & 1098 \\ \hline
 $P_{5}$ & 2289 & \caja{2091\\ 2175} & \caja{1527\\ 1841}& \caja{1762\\ 1868}& \caja{825\\ 996} &\caja{216\\ 851}& \caja{205\\ 447; 845\\ 1471; 1475}& \caja{845\\ 1476} & \caja{97.4\\124.7\\ 1148} & \caja{306\\ 1148} & 1151 \\ \hline
 $P_{6}$ & 2585 & \caja{2358\\ 2455} & \caja{1728\\ 2064}& \caja{1986\\ 2095}& \caja{939\\ 1133} &\caja{178\\ 955}& \caja{167\\ 513; 967\\ 1653; 1657}& \caja{967\\ 1657} & \caja{98.5\\124.4\\ 1274} & \caja{223\\ 1274} & 1277 \\ \hline
 $P_{7}$ & 2663 & \caja{2428\\ 2528} & \caja{1809\\ 2134}& \caja{2046\\ 2161}& \caja{970\\ 1169} &\caja{204\\ 990}& \caja{193\\ 530; 999\\ 1712; 1716}& \caja{999\\ 1716} & \caja{93.9\\123.4\\ 1227} & \caja{287\\ 1227} & 1230 \\ \hline
 $P_{8}$ & 2769 & \caja{2525\\ 2629} & \caja{1862\\ 2207}& \caja{2127\\ 2238}& \caja{1011\\ 1219} &\caja{164\\ 1023}& \caja{153\\ 554; 1043\\ 1770; 1774}& \caja{1043\\ 1774} & \caja{97.9\\123.7\\ 1330} & \caja{192\\ 1329} & 1332 \\ \hline
\end{tabular}
\end{center} \caption{Supersymmetric spectrum and Higgs masses for the set of benchmark points. All the masses are given in GeV. \label{tab:spectrum}}
\end{table}

 \section{Discussion}
\label{sec:discuss}

The most elegant solution to the $\mu$-problem in the MSSM is perhaps its extension to the
scale invariant NMSSM model.  In the MSSM, the existence of unifying underlying symmetries suggest 
the existence of unified boundary conditions leading to simple structures like that of the 
constrained MSSM, the CMSSM,  with only a few free parameters,  i.e.  $m$, $M$, $A$, $\mu$ and $B$. In contrast, 
unlike the MSSM case, there is some ambiguity in the 
definition of what a constrained version of the NMSSM, the CNMSSM, could be.

In this paper we have analyzed in detail a constrained version of the NMSSM model with boundary conditions 
obtained from the assumption of modulus dominated SUSY breaking in F-theory $SU(5)$ unification
models.  Such theories provide for an ultraviolet   completion for more traditional $SU(5)$ GUTs 
and provide for new solutions for problems like  doublet-triplet splitting and  D-quark  to lepton mass ratios.
One obtains a simple and very predictive structure of SUSY breaking soft terms summarized in eqs.(\ref{fgut}).
Additional approximate boundary conditions eq.(\ref{lasotrassoft})   may also be derived for soft terms involving
the NMSSM singlet $S$. Furthermore, the geometrical structure of F-theory unification implies that the
NMSSM couplings $\lambda$ and $\kappa$ are small with $\lambda,\kappa \leq 0.1$.  In spite of all these very
constrained parameter values, we find that the obtained very constrained NMSSM model is consistent with
correct EW symmetry breaking, a $125$ GeV Higgs and appropriate relic dark matter. The model passes also
B-decay constraints as well as LEP and LHC limits. 

Independently of its underlying string theory motivation, the structure of soft terms
(\ref{fgut}) and (\ref{lasotrassoft}) provides for a  definition of a constrained NMSSM model
consistent with all presently available data. The resulting NMSSM model is very predictive and the low-energy
spectrum very constrained, as
it is visible from  table (\ref{tab:spectrum}). The second lightest Higgs scalar $H_2$ can get a mass 
$m_{H_2}\simeq 125$ GeV whereas the lightest scalar $H_1$, with a dominant singlet component,  barely escaped
detection at LEP and could be observable at LHC as a peak  in $H_1\rightarrow \gamma \gamma$ at around  100 GeV. 
The LSP is mostly singlino and may provide for the correct relict density thanks to coannihilation with the lightest stau, 
which is the NLSP with a mass in the range 100-250 GeV. Such stau NLSP leads to signals at LHC involving
multi-tau events with missing energy. In this model the value of tan~$\beta$ is very large, tan~$\beta \simeq 50$, still the branching ratio  
for $B_s\rightarrow \mu^+\mu^-$ is below recent LHCb and CMS bounds and in many cases is smaller than the SM prediction, 
due to an interference effect. Gluinos and squarks have masses in the $2-3$ TeV range with a stop with a mass as low as 
1.2 TeV, all within reach of LHC(14).  On the other hand no large enhancement of the  $H_2\rightarrow \gamma \gamma$ over that
of the SM is expected. It is exciting to think that, 
if indeed the signatures above are observed at LHC, it will be evidence not only for SUSY but for an underlying
F-theory unified structure.

\vspace{2.0cm}

\centerline{\bf \large Acknowledgments}

\bigskip

We thank U.~Ellwanger, F.~Marchesano and  E.~Palti  for useful discussions. We are grateful to  F.~Staub for providing us with the modified  \texttt{SARAH-Spheno} NMSSM implementation of \cite{Staub:2010ty}. 
This work has been partially supported by the grants FPA 2009-09017, FPA 2009-07908, FPA2009-08958, FPA 2010-20807, FPA2012-34694, Consolider-CPAN 
(CSD2007-00042) and MultiDark (CSD2009-00064) from the Spanish MICINN, HEPHACOS-S2009/ESP1473 from the C.A. de Madrid, AGAUR 2009-SGR-168 from the Generalitat de Catalunya and the contract ``UNILHC" PITN-GA-2009-237920 of the European Commission. The IFT authors also thank the 
spanish MINECO {\it Centro de excelencia Severo Ochoa Program} under grant SEV-2012-0249.
D.G.C. is supported by the MICINN Ram\'on y Cajal programme through the grant RYC-2009-05096.

\newpage

\appendix

\section{The Higgs sector at one loop}
\label{app}

For completeness, in this appendix we repeat the analysis of section \ref{subsec:higgs} taking only into account one loop radiative corrections, as well as the two loop contributions to the bottom and top Yukawa couplings to leading logarithmic approximation. Namely, in this appendix we do not consider two loop radiative corrections to the Higgs sector. As we summarize below, the results however agree qualitatively with the ones in section \ref{subsec:higgs}, showing the robustness of our conclusions.

\begin{figure}[!t]
\begin{center}
\vspace{0.3cm}
\includegraphics[width=15.2cm]{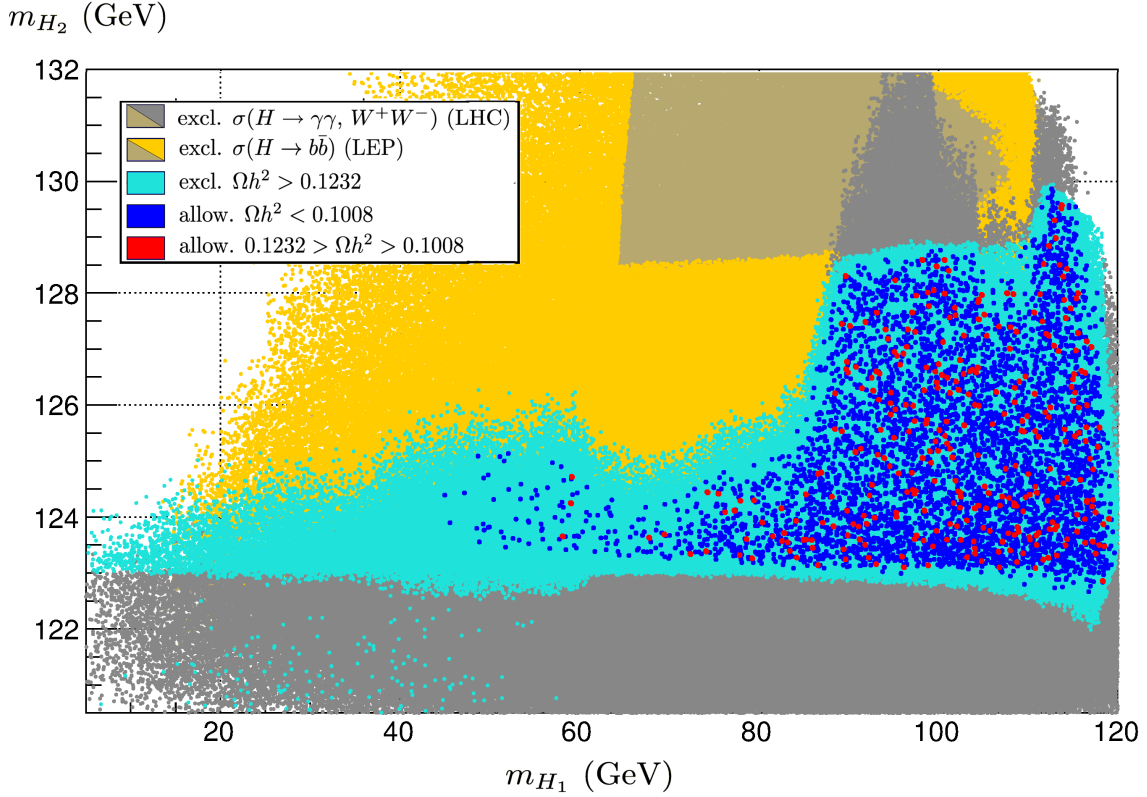}
\caption{\small Distribution of the light Higgs masses $(m_{H_1},m_{H_2})$ for NMSSM vacua with F-theory unification boundary conditions (\ref{fgut}) and unconstrained values of $A_\lambda$, $A_\kappa$ and $\lambda$, computed at one loop.\label{fig:h1h01loop}}
\end{center}
\end{figure}

\begin{figure}[!t]
\begin{center}
\vspace{0.3cm}
\includegraphics[width=15.2cm]{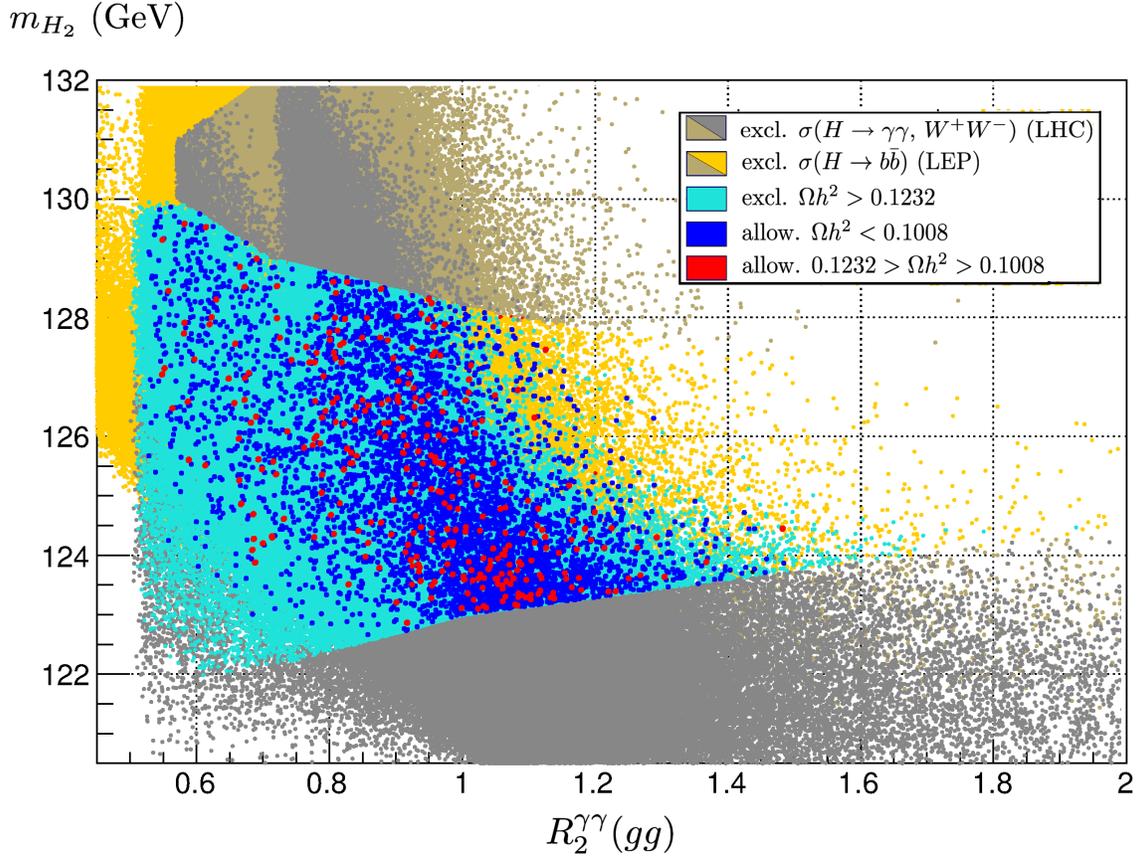}
\caption{\small Distribution of the the reduced signal cross section in the $H_2\to \gamma\gamma$ channel, $R_2^{\gamma\gamma}(gg\to H_2)$, versus the mass for the SM-like Higgs $H_2$ computed at one loop in NMSSM vacua with F-theory unification boundary conditions (\ref{fgut}) and unconstrained values of $A_\lambda$, $A_\kappa$ and $\lambda$.\label{fig:h1h1gg1loop}}
\end{center}
\end{figure}

In the models that we consider in this paper two loop radiative corrections tend to lower Higgs masses in as much as 2 - 5 GeV. Thus, at one loop we find in general higher masses than those in section \ref{subsec:higgs}. In spite of this, the mass of the lightest Higgs, $H_1$, is still not large enough to fit the LHC Higgs signal at any region of the parameter space, so that the role of the SM-like Higgs boson has again to be played by $H_2$. In figure \ref{fig:h1h01loop} we represent the distribution of masses $(m_{H_1},m_{H_2})$ for the two lightest Higgs bosons computed at one loop. One difference with respect to the two loop analysis performed in section \ref{subsec:higgs} is that the LHC bound on $\sigma^{W^+W^-}(gg\to H)$ now becomes also relevant, excluding solutions in the region $m_{H_1}\gtrsim 110$~GeV, $m_{H_2}\gtrsim 129$~GeV. Moreover, the LEP bound on $\sigma^{b\bar bZ}(e^+e^-\to HZ)$ and the LHC bound on $\sigma^{\gamma\gamma}(gg\to H)$ become softer, leading to the range of masses $50 \textrm{ GeV}\lesssim m_{H_1}\lesssim 120$~GeV and $123\textrm{ GeV}\lesssim m_{H_2}\lesssim 129$~GeV. The upper limit $m_{A_1}\lesssim 350$ GeV on the mass of the lightest pseudoscalar, on the other hand, still holds.

Regarding the reduced cross sections of the SM-like Higgs boson $H_2$, we find that the contribution from stau loops to the diphoton rate production become relevant in some cases, allowing for small enhancements with respect to the SM, as it can be seen in figure \ref{fig:h1h1gg1loop}. More precisely,  we observe from that figure that the reduced signal cross section $R_2^{\gamma\gamma}(gg\to H_2)$ lays in the range $0.6 - 1.4$.


\newpage

\begin{thebibliography}{99}

\bibitem{:2012gu}
  S.~Chatrchyan {\it et al.}  [CMS Collaboration],
  ``Observation of a new boson at a mass of 125 GeV with the CMS experiment at the LHC,''
  Phys.\ Lett.\ B {\bf 716} (2012) 30
  [arXiv:1207.7235 [hep-ex]].

\bibitem{:2012gk}
  G.~Aad {\it et al.}  [ATLAS Collaboration],
  ``Observation of a new particle in the search for the Standard Model Higgs boson with the ATLAS detector at the LHC,''
  Phys.\ Lett.\ B {\bf 716} (2012) 1
  [arXiv:1207.7214 [hep-ex]].

\bibitem{Maniatis:2009re}
  M.~Maniatis,
  ``The Next-to-Minimal Supersymmetric extension of the Standard Model reviewed,''
  Int.\ J.\ Mod.\ Phys.\ A {\bf 25} (2010) 3505
  [arXiv:0906.0777 [hep-ph]].

\bibitem{Ellwanger:2009dp}
  U.~Ellwanger, C.~Hugonie and A.~M.~Teixeira,
  ``The Next-to-Minimal Supersymmetric Standard Model,''
  Phys.\ Rept.\  {\bf 496} (2010) 1
  [arXiv:0910.1785 [hep-ph]].

\bibitem{thebook}
L.E. Ib\'a\~nez and A. Uranga,
``String Theory and Partiicle Physics: 
An Introduction to String Phenomenology.''
Cambridge Univiersity Press (2012).
  
\bibitem{Djouadi:2008uj}
  A.~Djouadi, U.~Ellwanger and A.~M.~Teixeira,
  ``Phenomenology of the constrained NMSSM,''
  JHEP {\bf 0904} (2009) 031
  [arXiv:0811.2699 [hep-ph]].


\bibitem{Djouadi:2008yj}
  A.~Djouadi, U.~Ellwanger and A.~M.~Teixeira,
  ``The Constrained next-to-minimal supersymmetric standard model,''
  Phys.\ Rev.\ Lett.\  {\bf 101} (2008) 101802
  [arXiv:0803.0253 [hep-ph]].


\bibitem{Ellwanger:2011sk}
  U.~Ellwanger,
  ``Higgs Bosons in the Next-to-Minimal Supersymmetric Standard Model at the LHC,''
  Eur.\ Phys.\ J.\ C {\bf 71} (2011) 1782
  [arXiv:1108.0157 [hep-ph]].


\bibitem{Hall:2011aa}
  L.~J.~Hall, D.~Pinner and J.~T.~Ruderman,
  ``A Natural SUSY Higgs Near 126 GeV,''
  JHEP {\bf 1204} (2012) 131
  [arXiv:1112.2703 [hep-ph]].

\bibitem{Ellwanger:2011aa}
  U.~Ellwanger,
  ``A Higgs boson near 125 GeV with enhanced di-photon signal in the NMSSM,''
  JHEP {\bf 1203} (2012) 044
  [arXiv:1112.3548 [hep-ph]].

\bibitem{Gunion:2012zd}
  J.~F.~Gunion, Y.~Jiang and S.~Kraml,
  ``The Constrained NMSSM and Higgs near 125 GeV,''
  Phys.\ Lett.\ B {\bf 710} (2012) 454
  [arXiv:1201.0982 [hep-ph]].

\bibitem{Arvanitaki:2011ck}
  A.~Arvanitaki and G.~Villadoro,
  ``A Non Standard Model Higgs at the LHC as a Sign of Naturalness,''
  JHEP {\bf 1202} (2012) 144
  [arXiv:1112.4835 [hep-ph]].

\bibitem{King:2012is}
  S.~F.~King, M.~Muhlleitner and R.~Nevzorov,
  ``NMSSM Higgs Benchmarks Near 125 GeV,''
  Nucl.\ Phys.\ B {\bf 860} (2012) 207
  [arXiv:1201.2671 [hep-ph]].
  
\bibitem{Kang:2012sy}
  Z.~Kang, J.~Li and T.~Li,
  ``On Naturalness of the MSSM and NMSSM,''
  JHEP {\bf 1211} (2012) 024
  [arXiv:1201.5305 [hep-ph]].

\bibitem{Cao:2012fz}
  J.~-J.~Cao, Z.~-X.~Heng, J.~M.~Yang, Y.~-M.~Zhang and J.~-Y.~Zhu,
  ``A SM-like Higgs near 125 GeV in low energy SUSY: a comparative study for MSSM and NMSSM,''
  JHEP {\bf 1203} (2012) 086
  [arXiv:1202.5821 [hep-ph]].

\bibitem{Ellwanger:2012ke}
  U.~Ellwanger and C.~Hugonie,
  ``Higgs bosons near 125 GeV in the NMSSM with constraints at the GUT scale,''
  Adv.\ High Energy Phys.\  {\bf 2012} (2012) 625389
  [arXiv:1203.5048 [hep-ph]].

\bibitem{Benbrik:2012rm}
  R.~Benbrik, M.~Gomez Bock, S.~Heinemeyer, O.~Stal, G.~Weiglein and L.~Zeune,
  ``Confronting the MSSM and the NMSSM with the Discovery of a Signal in the two Photon Channel at the LHC,''
  Eur.\ Phys.\ J.\ C {\bf 72} (2012) 2171
  [arXiv:1207.1096 [hep-ph]].

\bibitem{Gunion:2012gc}
  J.~F.~Gunion, Y.~Jiang and S.~Kraml,
  ``Could two NMSSM Higgs bosons be present near 125 GeV?,''
  Phys.\ Rev.\ D {\bf 86} (2012) 071702
  [arXiv:1207.1545 [hep-ph]].

\bibitem{Cao:2012yn}
  J.~Cao, Z.~Heng, J.~M.~Yang and J.~Zhu,
  ``Status of low energy SUSY models confronted with the LHC 125 GeV Higgs data,''
  JHEP {\bf 1210} (2012) 079
  [arXiv:1207.3698 [hep-ph]].

\bibitem{Belanger:2012tt}
  G.~Belanger, U.~Ellwanger, J.~F.~Gunion, Y.~Jiang, S.~Kraml and J.~H.~Schwarz,
  ``Higgs Bosons at 98 and 125 GeV at LEP and the LHC,''
  arXiv:1210.1976 [hep-ph].

\bibitem{Kowalska:2012gs}
  K.~Kowalska, S.~Munir, L.~Roszkowski, E.~M.~Sessolo, S.~Trojanowski and Y.~-L.~S.~Tsai,
  ``The Constrained NMSSM with a 125 GeV Higgs boson -- A global analysis,''
  arXiv:1211.1693 [hep-ph].

\bibitem{King:2012tr}
  S.~F.~King, M.~Muhlleitner, R.~Nevzorov and K.~Walz,
  ``Natural NMSSM Higgs Bosons,''
  arXiv:1211.5074 [hep-ph].
  
  \bibitem{ftheoryreviews}
    J.~J.~Heckman,
  { ``Particle Physics Implications of F-theory,''}
  arXiv:1001.0577 [hep-th]\\
    T.~Weigand,
  { ``Lectures on F-theory compactifications and model building,''}
  Class.\ Quant.\ Grav.\  {\bf 27}, 214004 (2010)
  [arXiv:1009.3497 [hep-th]]\\
    L.~E.~Ib\'a\~nez,
  ``From Strings to the LHC: Les Houches Lectures on String Phenomenology,''
  arXiv:1204.5296 [hep-th]\\
  G.~K.~Leontaris,
  ``Aspects of F-Theory GUTs,''
  PoS CORFU {\bf 2011} (2011) 095
  [arXiv:1203.6277 [hep-th]]\\
  A.~Maharana and E.~Palti,
  ``Models of Particle Physics from Type IIB String Theory and F-theory: A Review,''
  arXiv:1212.0555 [hep-th].
  
  

\bibitem{Aparicio:2008wh}
  L.~Aparicio, D.~G.~Cerdeno and L.~E.~Ibanez,
  ``Modulus-dominated SUSY-breaking soft terms in F-theory and their test at LHC,''
  JHEP {\bf 0807} (2008) 099
  [arXiv:0805.2943 [hep-ph]].

\bibitem{Aparicio:2012iw}
  L.~Aparicio, D.~G.~Cerdeno and L.~E.~Ibanez,
  ``A 119-125 GeV Higgs from a string derived slice of the CMSSM,''
  JHEP {\bf 1204} (2012) 126
  [arXiv:1202.0822 [hep-ph]].
  
\bibitem{Aaij:2012ac}
  R.~Aaij {\it et al.}  [LHCb Collaboration],
  ``Strong constraints on the rare decays $B_s \to \mu^+ \mu^-$ and $B^0 \to \mu^+ \mu^-$,''
  Phys.\ Rev.\ Lett.\  {\bf 108} (2012) 231801
  [arXiv:1203.4493 [hep-ex]].

\bibitem{Martini:2012np}
  L.~Martini [CMS Collaboration],
  ``Search for $B_s -> \mu^+ \mu^-$ and $B_0 -> \mu^+ \mu^-$ decays in CMS,''
  EPJ Web Conf.\  {\bf 28} (2012) 12038
  [arXiv:1201.4257 [hep-ex]].
  
  
\bibitem{mumucomb}
  The ATLAS, CMS and LHCb Collaborations,
  ``Search for the rare decays B-> mumu at the LHC with the ATLAS, CMS and LHCb experiments,''
  ATLAS-CONF-2012-061, CMS-PAS-BPH-12-009, LHCb-CONF-2012-017.
  
\bibitem{:2012ct}
  Aaij {\it et al.}  [LHCb Collaboration],
  ``First evidence for the decay Bs -> mu+ mu-,''
  arXiv:1211.2674 [Unknown].
  
\bibitem{Degrassi:2009yq}
  G.~Degrassi and P.~Slavich,
  ``On the radiative corrections to the neutral Higgs boson masses in the NMSSM,''
  Nucl.\ Phys.\ B {\bf 825} (2010) 119
  [arXiv:0907.4682 [hep-ph]].

  
  
\bibitem{ftheoryguts}
  R.~Donagi, M.~Wijnholt,
  { ``Model Building with F-Theory,''}
  [arXiv:0802.2969 [hep-th]]\\
  C.~Beasley, J.~J.~Heckman and C.~Vafa,
 { ``GUTs and Exceptional Branes in F-theory - I,''}
  JHEP {\bf 0901} (2009) 058
  [arXiv:0802.3391 [hep-th]].\\
  C.~Beasley, J.~J.~Heckman and C.~Vafa,
 { ``GUTs and Exceptional Branes in F-theory - II: Experimental Predictions,''}
  JHEP {\bf 0901} (2009) 059
  [arXiv:0806.0102 [hep-th]]\\
  R.~Donagi and M.~Wijnholt,
  { ``Breaking GUT Groups in F-Theory,''}
  arXiv:0808.2223 [hep-th].
 
 \bibitem{Aparicio:2012ju}
  L.~Aparicio,
  ``Some phenomenological aspects of Type IIB/F-theory string compactifications,''
  arXiv:1210.0339 [hep-th].
  

\bibitem{Font:2012wq}
  A.~Font, L.~E.~Ibanez, F.~Marchesano and D.~Regalado,
  ``Non-perturbative effects and Yukawa hierarchies in F-theory SU(5) Unification,''
  arXiv:1211.6529 [hep-th]\\
   L.~Aparicio, A.~Font, L.~E.~Ibanez and F.~Marchesano,
  ``Flux and Instanton Effects in Local F-theory Models and Hierarchical Fermion Masses,''
  JHEP {\bf 1108} (2011) 152
  [arXiv:1104.2609 [hep-th]].
 
\bibitem{Marsano:2009wr}
  J.~Marsano, N.~Saulina and S.~Schafer-Nameki,
  ``Compact F-theory GUTs with U(1) (PQ),''
  JHEP {\bf 1004} (2010) 095
  [arXiv:0912.0272 [hep-th]].

\bibitem{Dudas:2010zb}
  E.~Dudas and E.~Palti,
  ``On hypercharge flux and exotics in F-theory GUTs,''
  JHEP {\bf 1009} (2010) 013
  [arXiv:1007.1297 [hep-ph]].
  
\bibitem{Marsano:2010sq}
  J.~Marsano,
  ``Hypercharge Flux, Exotics, and Anomaly Cancellation in F-theory GUTs,''
  Phys.\ Rev.\ Lett.\  {\bf 106} (2011) 081601
  [arXiv:1011.2212 [hep-th]].
  
\bibitem{Palti:2012dd}
  E.~Palti,
  ``A Note on Hypercharge Flux, Anomalies, and U(1)s in F-theory GUTs,''
  arXiv:1209.4421 [hep-th].


\bibitem{Ludeling:2011en}
  C.~Ludeling, H.~P.~Nilles and C.~C.~Stephan,
  ``The Potential Fate of Local Model Building,''
  Phys.\ Rev.\ D {\bf 83} (2011) 086008
  [arXiv:1101.3346 [hep-th]].
  
\bibitem{swisscheese}
  V.~Balasubramanian, P.~Berglund, J.~P.~Conlon and F.~Quevedo,
  ``Systematics of moduli stabilisation in Calabi-Yau flux compactifications,''
  JHEP {\bf 0503} (2005) 007
  [hep-th/0502058]\\
  J.~P.~Conlon, F.~Quevedo and K.~Suruliz,
  ``Large-volume flux compactifications: Moduli spectrum and D3/D7 soft supersymmetry breaking,''
  JHEP {\bf 0508} (2005) 007
  [hep-th/0505076]\\
   M.~Cicoli, J.~P.~Conlon and F.~Quevedo,
  ``General Analysis of LARGE Volume Scenarios with String Loop Moduli Stabilisation,''
  JHEP {\bf 0810} (2008) 105
  [arXiv:0805.1029 [hep-th]]\\
   M.~Cicoli, C.~P.~Burgess and F.~Quevedo,
  ``Anisotropic Modulus Stabilisation: Strings at LHC Scales with Micron-sized Extra Dimensions,''
  JHEP {\bf 1110} (2011) 119
  [arXiv:1105.2107 [hep-th]].


\bibitem{Conlon:2006tj}
  J.~P.~Conlon, D.~Cremades and F.~Quevedo,
  ``Kahler potentials of chiral matter fields for Calabi-Yau string compactifications,''
  JHEP {\bf 0701} (2007) 022
  [hep-th/0609180].  
  
  
   
\bibitem{Cvetic:2010dz}
  M.~Cvetic, J.~Halverson and P.~Langacker,
  ``Singlet Extensions of the MSSM in the Quiver Landscape,''
  JHEP {\bf 1009} (2010) 076
  [arXiv:1006.3341 [hep-th]].
    
\bibitem{berasaluce}
  M.~Berasaluce-Gonzalez, L.~E.~Ibanez, P.~Soler and A.~M.~Uranga,
  ``Discrete gauge symmetries in D-brane models,''
  JHEP {\bf 1112} (2011) 113
  [arXiv:1106.4169 [hep-th]].
  
  
\bibitem{softfromflux}
  P.~G.~C\'amara, L.~E.~Ib\'a\~nez and A.~M.~Uranga,
  ``Flux induced SUSY breaking soft terms,''
  Nucl.\ Phys.\ B {\bf 689} (2004) 195
  [hep-th/0311241];\\
  D.~Lust, S.~Reffert and S.~Stieberger,
  ``Flux-induced soft supersymmetry breaking in chiral type IIB orientifolds with D3 / D7-branes,''
  Nucl.\ Phys.\ B {\bf 706} (2005) 3
  [hep-th/0406092];\\
  P.~G.~C\'amara, L.~E.~Ib\'a\~nez and A.~M.~Uranga,
  ``Flux-induced SUSY-breaking soft terms on D7-D3 brane systems,''
  Nucl.\ Phys.\  B {\bf 708} (2005) 268
  [arXiv:hep-th/0408036];\\
   D.~Lust, S.~Reffert and S.~Stieberger,
  ``MSSM with soft SUSY breaking terms from D7-branes with fluxes,''
  Nucl.\ Phys.\ B {\bf 727} (2005) 264
  [hep-th/0410074];\\
 M.~Grana, T.~W.~Grimm, H.~Jockers and J.~Louis,
  ``Soft supersymmetry breaking in Calabi-Yau orientifolds with D-branes and fluxes,''
  Nucl.\ Phys.\ B {\bf 690} (2004) 21
  [hep-th/0312232].
 

\bibitem{Beringer:1900zz}
  J.~Beringer {\it et al.}  [Particle Data Group Collaboration],
  ``Review of Particle Physics (RPP),''
  Phys.\ Rev.\ D {\bf 86} (2012) 010001.


\bibitem{Ellwanger:2004xm}
  U.~Ellwanger, J.~F.~Gunion and C.~Hugonie,
  ``NMHDECAY: A Fortran code for the Higgs masses, couplings and decay widths in the NMSSM,''
  JHEP {\bf 0502} (2005) 066
  [hep-ph/0406215];
  
  U.~Ellwanger and C.~Hugonie,
  ``NMHDECAY 2.0: An Updated program for sparticle masses, Higgs masses, couplings and decay widths in the NMSSM,''
  Comput.\ Phys.\ Commun.\  {\bf 175} (2006) 290
  [hep-ph/0508022];
  
  U.~Ellwanger and C.~Hugonie,
  ``NMSPEC: A Fortran code for the sparticle and Higgs masses in the NMSSM with GUT scale boundary conditions,''
  Comput.\ Phys.\ Commun.\  {\bf 177} (2007) 399
  [hep-ph/0612134].
  
\bibitem{Aaltonen:2012ra}
  T.~Aaltonen {\it et al.}  [CDF and D0 Collaborations],
  ``Combination of the top-quark mass measurements from the Tevatron collider,''
  Phys.\ Rev.\ D {\bf 86} (2012) 092003
  [arXiv:1207.1069 [hep-ex]].
  
\bibitem{htautau}
  The CMS Collaboration,
  ``Search for the standard model Higgs boson decaying to tau pairs,''
  CMS-PAS-HIG-12-043.

  
\bibitem{hww}
  The CMS Collaboration,
  ``Evidence for a particle decaying to $W^+W^-$ in the fully leptonic final state in a standard model Higgs boson search in $pp$ collisions at the LHC,''
  CMS-PAS-HIG-12-042.
  
\bibitem{hzz}
  The CMS Collaboration,
  ``Updated results on the new boson discovered in the search for the standard model Higgs boson in the $H \to ZZ \to 4\ell$ channel in $pp$ collisions at $\sqrt{s} =$ 7 and 8 TeV,''
  CMS-PAS-HIG-12-041.
  
\bibitem{hbb}
  The CMS Collaboration,
  ``Search for the standard model Higgs boson produced in association with W or Z bosons, and decaying to bottom quarks,''
  CMS-PAS-HIG-12-043.  

\bibitem{cmsgamma}
  The CMS Collaboration,
  ``Evidence for a new state decaying into two photons in the search for the standard model Higgs boson in pp collisions,''
  CMS-PAS-HIG-12-015.

\bibitem{atlasgamma}
  The ATLAS Collaboration,
  ``Observation of an excess of events in the search for the Standard Model Higgs boson in the gamma-gamma channel with the ATLAS detector,''
  ATLAS-CONF-2012-091.

\bibitem{Aad:2012tj}
  G.~Aad {\it et al.}  [ATLAS Collaboration],
  ``Search for charged Higgs bosons decaying via $H^{+} \to \tau \nu$ in top quark pair events using $pp$ collision data at $\sqrt{s}=7$ TeV with the ATLAS detector,''
  JHEP {\bf 1206} (2012) 039
  [arXiv:1204.2760 [hep-ex]].
  
\bibitem{Belanger:2001fz}
  G.~Belanger, F.~Boudjema, A.~Pukhov and A.~Semenov,
  ``micrOMEGAs: A program for calculating the relic density in the MSSM,''
  Comput.\ Phys.\ Commun.\  {\bf 149} (2002) 103
  [arXiv:hep-ph/0112278];

  G.~Belanger, F.~Boudjema, A.~Pukhov and A.~Semenov,
  ``micrOMEGAs: Version 1.3,''
  Comput.\ Phys.\ Commun.\  {\bf 174}, 577 (2006)
  [arXiv:hep-ph/0405253];

  G.~Belanger, F.~Boudjema, P.~Brun, A.~Pukhov, S.~Rosier-Lees, P.~Salati and A.~Semenov,
  ``Indirect search for dark matter with micrOMEGAs2.4,''
  Comput.\ Phys.\ Commun.\  {\bf 182} (2011) 842
  [arXiv:1004.1092 [hep-ph]].
  
  
\bibitem{Komatsu:2010fb}
  E.~Komatsu {\it et al.}  [WMAP Collaboration],
  ``Seven-Year Wilkinson Microwave Anisotropy Probe (WMAP) Observations: Cosmological Interpretation,''
  Astrophys.\ J.\ Suppl.\  {\bf 192} (2011) 18
  [arXiv:1001.4538 [astro-ph.CO]].
  
  

  
 
\bibitem{Ross:2007az}
  G.~Ross and M.~Serna,
  ``Unification and fermion mass structure,''
  Phys.\ Lett.\ B {\bf 664} (2008) 97
  [arXiv:0704.1248 [hep-ph]].

\bibitem{Elor:2012ig}
  G.~Elor, L.~J.~Hall, D.~Pinner and J.~T.~Ruderman,
  ``Yukawa Unification and the Superpartner Mass Scale,''
  JHEP {\bf 1210} (2012) 111
  [arXiv:1206.5301 [hep-ph]].

\bibitem{Staub:2008uz}
  F.~Staub,
  ``Sarah,''
  arXiv:0806.0538 [hep-ph];

  F.~Staub,
  ``From Superpotential to Model Files for FeynArts and CalcHep/CompHep,''
  Comput.\ Phys.\ Commun.\  {\bf 181} (2010) 1077
  [arXiv:0909.2863 [hep-ph]];

  F.~Staub,
  ``Automatic Calculation of supersymmetric Renormalization Group Equations and Self Energies,''
  Comput.\ Phys.\ Commun.\  {\bf 182} (2011) 808
  [arXiv:1002.0840 [hep-ph]].

\bibitem{Staub:2010ty}
  F.~Staub, W.~Porod and B.~Herrmann,
  ``The Electroweak sector of the NMSSM at the one-loop level,''
  JHEP {\bf 1010} (2010) 040
  [arXiv:1007.4049 [hep-ph]].

\bibitem{Barate:2003sz}
  R.~Barate {\it et al.}  [LEP Working Group for Higgs boson searches and ALEPH and DELPHI and L3 and OPAL Collaborations],
  ``Search for the standard model Higgs boson at LEP,''
  Phys.\ Lett.\ B {\bf 565} (2003) 61
  [hep-ex/0306033].

\bibitem{Cerdeno:2004xw}
  D.~G.~Cerde\~no, C.~Hugonie, D.~E.~Lopez-Fogliani, C.~Mu\~noz and
  A.~M.~Teixeira, 
  ``Theoretical predictions for the direct detection of neutralino dark  matter
  in the NMSSM,''
  JHEP {\bf 0412} (2004) 048
  [arXiv:hep-ph/0408102]. 

\bibitem{Buras:2002vd}
  A.~J.~Buras, P.~H.~Chankowski, J.~Rosiek and L.~Slawianowska,
  ``$\Delta M_{d,s}, B^0{d,s} \to \mu^{+} \mu^{-}$ and $B \to X_{s} \gamma$ in supersymmetry at large $\tan\beta$,''
  Nucl.\ Phys.\ B {\bf 659} (2003) 3
  [hep-ph/0210145].

\bibitem{Bobeth:2001jm}
  C.~Bobeth, A.~J.~Buras, F.~Kruger and J.~Urban,
  ``QCD corrections to $\bar{B} \to X_{d,s} \nu \bar{\nu}$, $\bar{B}_{d,s} \to \ell^{+} \ell^{-}$, $K \to \pi \nu \bar{\nu}$ and $K_{L} \to \mu^{+} \mu^{-}$ in the MSSM,''
  Nucl.\ Phys.\ B {\bf 630} (2002) 87
  [hep-ph/0112305].

\bibitem{Hiller:2004ii}
  G.~Hiller,
  ``B physics signals of the lightest CP odd Higgs in the NMSSM at large tan beta,''
  Phys.\ Rev.\ D {\bf 70} (2004) 034018
  [hep-ph/0404220].
  
\bibitem{Domingo:2007dx}
  F.~Domingo and U.~Ellwanger,
  ``Updated Constraints from $B$ Physics on the MSSM and the NMSSM,''
  JHEP {\bf 0712} (2007) 090
  [arXiv:0710.3714 [hep-ph]].

\bibitem{Hodgkinson:2008qk}
  R.~N.~Hodgkinson and A.~Pilaftsis,
  ``Supersymmetric Higgs Singlet Effects on B-Meson FCNC Observables at Large tan(beta),''
  Phys.\ Rev.\ D {\bf 78} (2008) 075004
  [arXiv:0807.4167 [hep-ph]].

\bibitem{Cerdeno:2007sn}
  D.~G.~Cerde\~no, E.~Gabrielli, D.~E.~Lopez-Fogliani, C.~Mu\~noz and
  A.~M.~Teixeira, 
  ``Phenomenological viability of neutralino dark matter in the NMSSM,''
  JCAP {\bf 0706} (2007) 008
  [arXiv:hep-ph/0701271].  

\bibitem{g-2}
  G.~W.~Bennett {\it et al.} [ Muon G-2 Collaboration ],
  ``Final Report of the Muon E821 Anomalous Magnetic Moment Measurement at BNL,''
  Phys.\ Rev.\  {\bf D73 } (2006)  072003.
  [hep-ex/0602035].

\bibitem{Hagiwara:2011af}
  K.~Hagiwara, R.~Liao, A.~D.~Martin, D.~Nomura and T.~Teubner,
  ``$(g-2)_\mu$ and $\alpha(M_{Z^2})$ re-evaluated using new precise data,''
  J.\ Phys.\ G G {\bf 38} (2011) 085003
  [arXiv:1105.3149 [hep-ph]].


\bibitem{Jegerlehner:2009ry}
  F.~Jegerlehner and A.~Nyffeler,
  ``The Muon g-2,''
  Phys.\ Rept.\  {\bf 477} (2009) 1
  [arXiv:0902.3360 [hep-ph]].
  
  \bibitem{Davier:2010nc}
  M.~Davier, A.~Hoecker, B.~Malaescu and Z.~Zhang,
  ``Reevaluation of the Hadronic Contributions to the Muon g-2 and to alpha(MZ),''
  Eur.\ Phys.\ J.\ C {\bf 71} (2011) 1515
  [arXiv:1010.4180 [hep-ph]].
  
\bibitem{htautauatlas}
  The ATLAS Collaboration, 
  ``Search for the Standard Model Higgs boson in $H\to \tau +\tau^-$ decays in proton-proton collisions with the ATLAS detector,'' 
  ATLAS-CONF-2012-160.    
  
  
\bibitem{Ellwanger:2010es}
  U.~Ellwanger, A.~Florent and D.~Zerwas,
  ``Discovering the constrained NMSSM with tau leptons at the LHC,''
  JHEP {\bf 1101} (2011) 103
  [arXiv:1011.0931 [hep-ph]].

\bibitem{Belanger:2012jn} 
  G.~Belanger, S.~Biswas, C.~Boehm and B.~Mukopadyaya,
  ``Light Neutralino Dark Matter in the MSSM and Its Implication for LHC Searches for Staus,''
  arXiv:1206.5404 [hep-ph].

  



\end{thebibliography}
\end{document}